\documentclass[preprint,12pt]{elsarticle}

\usepackage{amssymb}
\usepackage{graphicx,color}
\usepackage{epsfig}
\usepackage{dcolumn}
\usepackage{bm}

\usepackage{pstricks}

\newcommand{\grad}{{ \color{black} \vec \nabla}}

\newcommand{\etaeff}{{\color{black} \eta_{eff}}}

\newcommand{\GP}{ { \color{black}  G} }

\newcommand{\rug}{ {\color{black} \zeta}}  
\newcommand{\std}{ {\color{black} \sigma}}

\newcommand{\flowpower}{ {\color{black} \alpha}}
\newcommand{\Scutpower}{ {\color{black} \gamma}}
\newcommand{\Spower}{ {\color{black} \tau}}

\newcommand{\corrsize}{\color{black} \lambda}

\newcommand{\aScut}{ {\color{black} \tilde{S_0}}}
\newcommand{\aPcsyst}{ {\color{black} \Delta \tilde{P_c}}}
\newcommand{\alength}{  { \color{black} \tilde{L}}} 
\newcommand{\awidth}{  { \color{black} \tilde{W}}} 
\newcommand{\aGP}{ { \color{black}  \tilde{ \GP} }}

\newcommand{\eqref}[1]{(\ref{#1})}

\begin{document}

\begin{frontmatter}

\title{Numerical study of Bingham flow in macrosopic two dimensional heterogenous porous media}

\author{R. Kostenko}
\address{Laboratoire FAST, Univ Paris-Sud, CNRS, Universit\'e Paris-Saclay, F-91405, Orsay, France.\\ANDRA, 1/7 rue Jean Monnet, Parc de la Croix-Blanche, 92298 Ch{\^a}tenay-Malabry cedex, France}
\author{L. Talon}
\address{Laboratoire FAST, Univ Paris-Sud, CNRS, Universit\'e Paris-Saclay, F-91405, Orsay, France.}

\date{\today\ -- \jobname}

\begin{abstract} 

The flow of non-Newtonian fluids is ubiquitous in many applications in the geological and industrial context. We focus here on yield stress fluids (YSF), i.e. a material that requires minimal stress to flow.
We study numerically the flow of yield stress fluids in 2D porous media on a macroscopic scale in the presence of local heterogeneities. 
As with the microscopic problem, heterogeneities are of crucial importance because some regions will flow more easily than others.
As a result, the flow is characterized by preferential flow paths with fractal features.
These fractal properties are characterized by different scale exponents that will be  determined and analyzed.
One of the salient features of these results is that these exponents seem to be independent of the amplitude of heterogeneities for a log-normal distribution.
In addition, these exponents appear to differ from those at the microscopic level, illustrating the fact that, although similar, the two scales are governed by different sets of equations.

\end{abstract}

\end{frontmatter}

\section{\bf Introduction} 

The flow of non-Newtonian fluids is ubiquitous in many applications in geological and industrial context.
In this study, we focus on yield stress fluid (YSF), viz. material that requires a minimum amount of stress  to flow.
When the  stress is below a critical value, the material behave as solid. Once the stress is above this threshold, the material can be sheared and thus flow.

These materials have many implications for industrial processes. Indeed mud, polymers, oil, foam suspension may present a yielding threshold \cite{coussot05}.
 The most important is certainly the extraction of heavy oil and has been the subject of many studies since the 1960s \cite{entov67}. 
 Yield stress fluids are also used for enhanced oil recovery where foam or polymers are injected into the ground to block preferential flow paths \cite{prudhomme95,rossen90a}. 
 Another application is the hydraulic fracking where the yield stress properties is used to prevent the closing of the fractures \cite{barbati16}.

Because of the many applications, flow of yield stress fluids has been investigated in many studies \cite{entov67,al-fariss87,chase05,chen05,sochi08,chevalier13,talon13a,castro14,nash16,shahsavari16}.
Most of these studies focused on flow at the microscopic (pore) scale. The main objective was to establish a constitutive equation to relate the flow within the medium to the applied pressure drop, which represents a generalization of the Darcy's law to yield stress fluids.

Yield stress fluids are usually described by the Herschel-Bulkley model which relates the shear rate, $\dot \gamma$, to the applied shear stress $\tau$:
\begin{equation}
\dot \gamma = \left\{
 \begin{array}{lcc}
    0 & \rm{ if } & \tau<\tau_0 \\
	(\frac{1}{C}(\tau - \tau_0))^{1/n} & \rm{ if } & \tau>\tau_0\\
\end{array}
\right. ,
\end{equation}
where $ \tau_0$ is the yield stress, $C$ the consistency and $n$ an exponent.
Here, we choose to restrict to the Bingham rheology which corresponds to  $n=1$ (Bingham fluid) and $C=\eta_0$, the dynamic viscosity.

The usual governing equation for the flow of yield stress fluid at the field scale has been proposed by Entov \cite{entov67} and Pascal \cite{pascal81}.
They proposed a modified Darcy's equation (see also \cite{wu90,rees15}) in the form:
\begin{equation}
\label{eq:darcy_simple}
\vec u =  \left\{
\begin{array}{lcc}
  -\frac{k}{\eta_0} (\grad P - \GP \frac{\grad P}{|\grad P |}) = -\frac{k}{\eta_0} (\grad P + \GP \frac{\vec u}{|\vec u |})  & if & |\grad P| > \GP \\
    \vec 0 & if & |\grad P| < \GP
  \end{array}
  \right. ,
\end{equation}
where  $k$ is the permeability and $\GP$ a positive scalar representing the limiting pressure gradient related to the yield stress. Here, we have used the fact that the velocity vector and the gradient of pressure are opposed direction.
Since its introduction, this law has been validated by several experimental studies (see for instance \cite{al-fariss87,chase05,chevalier13}).
It is important to note that both the limiting pressure and the permeability depend on the local topology of the material. 
 These quantities may thus be subject to spacial variations at  the macroscopic geological scale.
 
In recent studies \cite{talon13a,chevalier15a,chevalier15c}, we have, however, demonstrated that, if the medium in sufficiently disordered, the flow curve might be non-linear  close to the limiting pressure.
Because of the structural disorder, some channel paths may be easier to flow than others. For a certain range of pressure, there is thus a non-linear increase of flowing paths responsible for the non-linearity of the flow rate curve.
At sufficiently high pressure, once all the fluid has yielded, the flow curve becomes linear again and  eq. \eqref{eq:darcy_simple} is valid.

One of the characteristics of these study was to demonstrate that the flow curve and structure were governed by a power law.
Indeed, the flow rate satisfies $Q \propto (\nabla P - \GP) ^{\alpha}$, where $\alpha$ is an exponent close to $2$.
Moreover, the size ($S$) distribution of the cluster of the fluid at rest follow : $P(S) \propto S^{-\tau} \exp{S/S_0}$ with $S_0 \propto Q^{-\gamma}$.
This distribution law characterizes the multi-scale aspect (fractal) of this problem. It is also reminiscent to other problems like  percolation \cite{stauffer91} or the avalanche dynamic \cite{amaral95,chevalier15c, chevalier17}.
It should also be noted that the exponent $\alpha$, $\tau$ and $\gamma$ were found to be independent of the type of disorder.

The aim of this study is to investigate the flow of a Bingham fluid at a  macroscopic level, i.e by solving a modified Darcy's law in a heterogenous permeability field.
As noted above, the permeability and the limiting pressure gradient depends on the local topology of the medium.
We therefore investigate the effect of spatial variations in these quantities on the flow.
For example, one could expect  some similarities with the microscale aspect such as the appearance of preferential paths. 
In their interesting work \cite{hewitt16}, Hewitt \emph{et al.} studied Bingham fluids in the Hele-Shaw cell in the presence of obstacles where they observed an increase in preferential paths with pressure in fractures (Hele-Shaw with open variations).

The article is divided as follows. First, we will present the problem, the flow equations and the description of the porous medium. Secondly, we will present the numerical methods based on a Lattice Boltzmann scheme with two relaxation times.
We will then present and discuss the results and the conclusion.

\section{Governing equations}

\subsection{Darcy's law}

The purpose of this paper is to study the flow of Bingham fluid at a macroscopic scale. To this goal, we solve the modified Darcy's equation which relates the local mean flow rate to the local mean gradient of pressure.

In the introduction, we have presented the standard Darcy's equation \eqref{eq:darcy_simple} proposed in the literature. 
As discussed, this equation is not true close to the limiting pressure gradient, where one  should expect a transitory regime which is more complex.
However, because the transition is not yet fully understood and because it appears in a short range of pressure, for the sake of simplicity, we will assume  that the standart eq. \eqref{eq:darcy_simple} is valid.

This equation can then be reformulated homogeneous to a balance of momentum:
\begin{equation}
\vec 0 = -\grad P - \frac{\eta_0}{k} \left( 1 + \frac{k \GP}{\eta_0 |u|}\right) \vec u.
\end{equation}
Here, we can see that the contribution of the yield stress can be seen as a constant body force opposed to the flow.

We have thus,
\begin{equation}
\vec 0 =  -\grad P - \frac{\etaeff(|u|)}{k} \vec u,
  \end{equation}
where $\etaeff(|u|) = \eta_0 \left( 1 + \frac{k \GP}{\eta_0 |u|}\right)$ is a scalar effective viscosity that depends on the local velocity field.
The model of effective viscosity is also very common to solve non-linear rheology in porous media.
Following the experimental work of Hirasaki and Pope \cite{hirasaki74} or the one of Chauveteau \cite{chauveteau82},  the idea is to introduce an effective shear rate related to the local mean velocity. The usual Darcy's law is then modified with an effective viscosity varying with the local 
mean velocity.

However, we have introduced two modifications to this equation.
First, like the Stokes equation for Bingham fluids, the effective viscosity has a zero velocity divergence point which can lead to numerical instabilities.
We have tackled this problem by capping the viscosity by a maximum value $\eta_{eff}^{max}$ which is several order of magnitude larger than $\eta_0$ (typically $10^9$ larger).
Another common technique to overcome this problem would have been to regularize this function around zero with an exponential term as proposed by Papanastasiou \cite{papanastasiou87}. We tried both methods which gave similar results but the second was  much slower.

The second comment is a general problem of the Darcy equation,  which is a first order differential equation, in disorder media.
Indeed, any discontinuity in the permeability field then leads to a discontinuity in the velocity field, which is not physical but also introduces numerical instabilites.
To overcome this problem, one possibility is to introduce a diffusive term into the momentum equation as proposed by Brinkman \cite{brinkman47}: 

\begin{equation}
\label{eq:Darcy_Brinkman}
\vec 0 = -\grad P  - \frac{\etaeff(|u|)}{k} \vec u +  \eta_B \Delta \vec u,
\end{equation}
where $\eta_B$ is a coefficient that depends on the viscosity and the geometry. Here, for the sake of simplicity, we keep it constant with $\eta_B = \eta_0$.

\subsection{Porous medium}

For the heterogenous permeability field, we use a log-normal distribution that has been widely used since the work of  Gehlar \cite{gelhar83} or Dagan \cite{dagan82} to describe heterogeneity at macroscopic scale.
The  permeabilty map is distributed according to: \begin{equation} \label{eq:distr_K}  pdf(f=\log k) = \frac{1}{\sqrt{2 \pi \sigma}}\exp{-\frac{(f-f_0)^2}{2 \sigma^2}}, \end{equation} where $f_0$ and $\sigma$ are the mean and the standard deviation of $\log(K)$.

The field $f=\log(k)$ is correlated with a Gaussian function, set by the correlation length $\lambda$:  \begin{equation} \label{eq:correlation}  \int f(\vec r+ \vec r_0)f(\vec r_0)) dxdy = e^{ -\frac{\pi}{8 \lambda^2} (x^2 + y^2) } \end{equation}
The field is generated using a standard Fast Fourier Method (see  Yiotis \emph{et al.} \cite{yiotis13} for details).
The permeability field $k(\vec r)$ is thus parametrized by three parameters, the mean $f_0$, the amplitude of the heterogeneities $\sigma$ and the correlation length $\lambda$. Moreover, we define $k_0=e^{f_0}$

From a phenomenological aspect, local critical pressure is expected to be related to permeability.
However, no clear relationship has been proposed in the literature. We have therefore chosen to relate them using dimensional arguments.
Indeed, one expects  that the critical pressure can be written as $\GP = \tau_0/d$, where $d$ is a geometrical factor that has the dimension of a length (the typical pore size) \cite{bird87}.
Since permeability has the dimension of a length to the power $2$, a natural relationship is to assume $\GP = A /\sqrt{k}$, where $A$ is a prefactor.
This relationship is, of course, purely phenomenological and cannot be true for any types of porous media.
We expect to hold for media with same topology such has mono disperse bead packing.

From the distribution of permeability eq. \eqref{eq:distr_K},   the probability density function of the limiting pressure also follows a log-normal distribution:

\begin{equation}
	p( \ln{\GP} ) = \sqrt{\frac{2}{\pi \sigma}} \exp{(-\frac{2 (\ln{\GP} - \ln{A/\sqrt{k_0}})^2}{\sigma^2})}.
\end{equation}

\subsection{Non-dimensionalization}

In the following, we will use dimensionless quantities.
For this, we nondimensionalize any  lengths with the correlation length, $\lambda$, any pressure with 
the average critical pressure gradient $A/\sqrt{k_0} \lambda$ and any velocities with the average permeability and the average critical pressure gradient :
$$X = \lambda \tilde X,  \; \; P = \lambda  A/\sqrt{k_0} \tilde P \; \; \rm{ and }  \; \; u = \frac{ A \sqrt{k_0} }{\eta_0} \tilde u, $$
where $\tilde{.}$ indicates dimensionless quantities.

\section{Numerical method}

To solve the Brinkman equation eq. \eqref{eq:Darcy_Brinkman}, we used an improved two-relaxation time Lattice Boltzman method (IBF-TRT) proposed by I. Ginzburg \cite{ginzburg15,ginzburg17}.
The basic idea of the Lattice Boltzmann method is to discretize the particle velocity distribution function  on a grid.
Here we used a 2-dimensional scheme with 9 different velocities ($\vec c_q$).
We then introduce the population ${f_q}$ as the density of particles moving with the velocity ${\vec c_q}$. 
The algorithm consists of a succession of two main steps. The first is the propagation step Eq. \eqref{trt2}, where we move the density on the grid according to its velocity.
The second is the collision step Eq. \eqref{trt_col}, where populations meeting at the same node are redistributed using a collision operator. This collision part is described as a relaxation toward an equilibrium state that depends on local macroscopic quantities (pressure, speed...).

The main idea of the TRT scheme is to decompose each population into a symmetrical and an antisymmetrical part.  Each component relax toward their equilibrium with its own  relaxation rate.
We define the even and odd part respectively as $f_q^+=\frac{f_q+f_{\bar{q}} }{2}$  and $f_q^-=\frac{f_q-f_{\bar{q}} }{2}$.
Denoting by $\bar q$ the direction opposite to $q$ ($\vec c_q = - \vec c_{\bar{•} q}$) and assuming  $\vec c_0=\vec 0$ and that the $8$ nonzero velocities are ordered so that the first four directions are opposite to the  last four. 

The propagation step is described as:
\begin{equation}
	\label{trt2}
	f_q(\vec{r}  + \vec c_q, t+1) = \hat f_q(\vec r, t),    q=0\dots 8,
\end{equation}
and the collision step:
\begin{eqnarray}
	\hat f_q(\vec r,t)= [ f_q - s^+ n_q^+ -  s^- n_q^-] , q=0\dots 4    \nonumber \\
	\hat f_{\bar q}  (\vec r,t)= [ f_{\bar q} - s^+ n_{ q}^+ +  s^- n_{ q}^-] ,  q=0\dots 4  ,
	\label{trt_col}
\end{eqnarray}
with
\begin{eqnarray}
\label{link}
{n_q^\pm}&=&({f_q^\pm}- e_q^{\pm})\;,\;   \;{\mbox {when}}\; \;{\vec c}_{\bar q}=-{\vec c_q}\;,\; q=0\ldots 4 \;,
\end{eqnarray}
where $e_q^{\pm}$ are the equilibrium distributions.
The model has several parameters, $c_s$ the numerical sound speed, $\nu_0$ the Brinkman viscosity and $\Lambda$ a numerical parameter, from which we define $\Lambda^+= 3 \eta_0$ and $\Lambda^-= \Lambda/ ( 3\eta_0)$.
The determination of the equilibrium function requires to compute the local pressure $P$ and momentum $\vec u$:

\begin{equation}
P = \rho c_s^2
  \;\; \rm{with} \;\; 
  \rho= \sum_0^8 f_q 
\end{equation}
and 

\begin{equation}
\vec u = \frac{2 \vec J}{2+\frac{\etaeff(\vec u)}{k}  }  \;\; \rm{with} \;\; \vec J = \sum_0^8 f_q \vec c_q,
\end{equation}
where the effective viscosity has be determined using the local velocity at the previous step.

The equilibrium function are then defined as:
\begin{equation}
e_q^+ = t_q P, \;  e_0 = \rho - \sum_1^{8} e_q^+
\end{equation}
and
\begin{equation}
e_q^- = t_q ( \vec u . \vec c_q + \Lambda^- F_q).
\end{equation} 
$F_q$ is the body force which takes into accound the Darcy drag force (second term in eq. \eqref{eq:Darcy_Brinkman}) and is defined by:
\begin{equation}
F_q = - t_q \frac{\etaeff(\vec u)}{k} \vec j. \vec c_q.
\end{equation}

Finally, the two relaxation rates are defined as:
\begin{equation}
	s^+ = \frac{2}{2 \tilde{\Lambda}^+ +1 } \;\; \rm{ with }
 \;\;
\tilde{\Lambda}^+ =  \frac{ 9 (4 + \frac{\etaeff(\vec u)}{\eta_0 k}) }{ 4 (3 + 2 \Lambda \frac{\etaeff(\vec u)}{\eta_0 k}) } \eta_0
\end{equation}
and 
\[
s^- = \frac{2}{2 {\Lambda}^- +1 }.
\]

The main difference between the standard BF-TRT and the improved IBF-TRT scheme is the use of $\tilde{\Lambda^+}$ instead of $\Lambda^+$ for the relaxation parameter $s^+$.
Indeed, the standard BF-TRT scheme leads to an error in the viscosity which depends on the local parameter. The change of $\Lambda^+$ is to compensate for this error \cite{ginzburg15,ginzburg17}.

In this study, the typical porous medium size is  $1024\times 4096$ nodes. The typical duration time is between  $2$ and $100$ hours with $32$ CPUs.

\subsection{Validation}

To validate our numerical code, we  simulated a Bingham fluid in a stratified porous medium with permeability distributed according to a sinusoidal function in the normal direction ($y$) and invariant in the flow direction ($x$), namely: $$\kappa(x,y) = \sin( 2 \pi y )\frac{\kappa_{max}-\kappa_{min}}{2} + 2\kappa_{min}.$$
We used different numbers of grid nodes, $N_y\in \{5,8,16,32,64,128\}$, to vary the spatial resolution.
To perform a  quantitative analysis,  we solve eq. \eqref{eq:Darcy_Brinkman} using a standard second order finite difference method discretized with a fixed finer resolution of $512$ points.

\begin{figure}
	\includegraphics[width=0.9\linewidth]{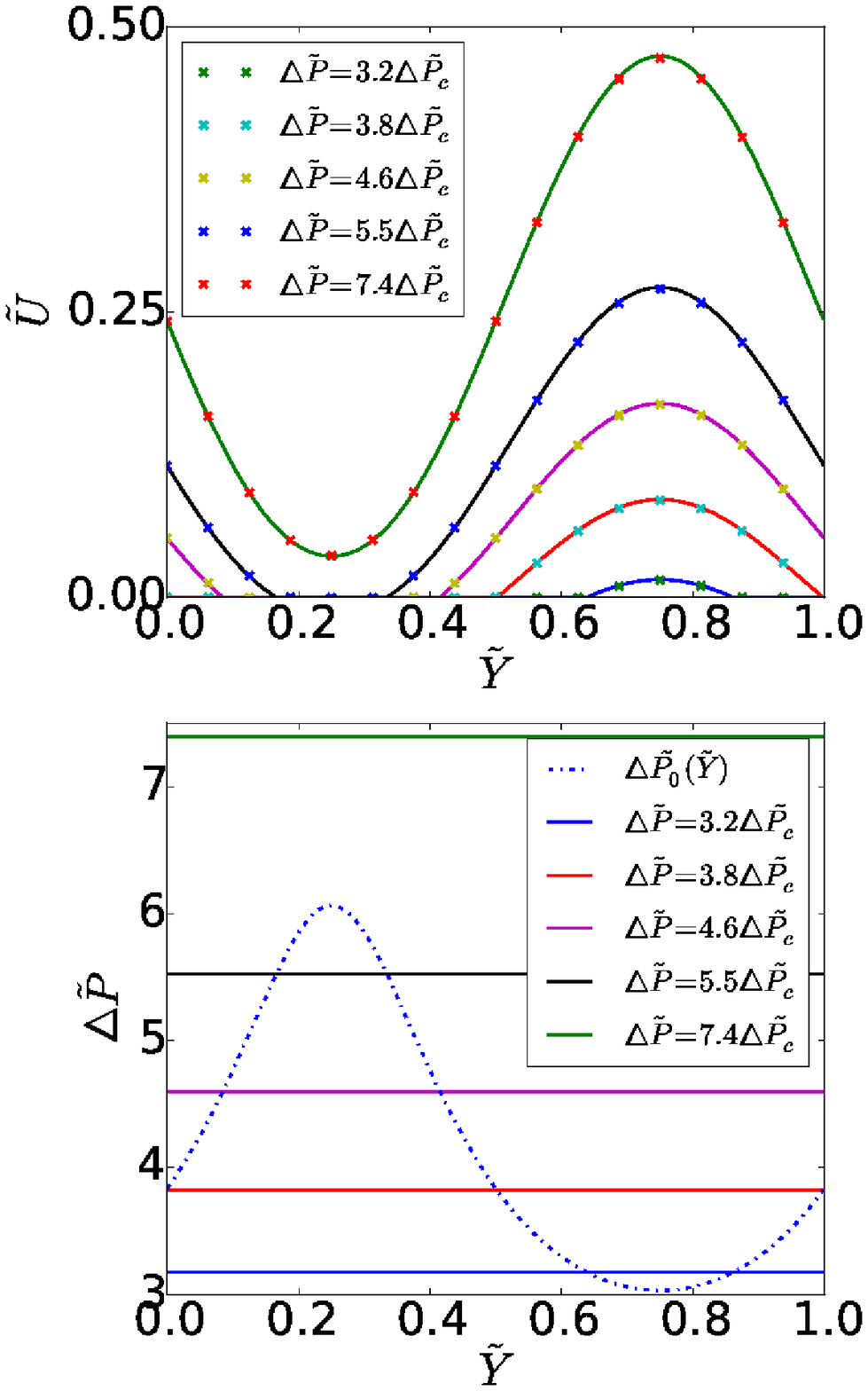}
	\caption{Top: Velocity profile $\tilde u( \tilde y)$ in a sinusoidal stratified permeability field with $(\kappa_{max};\kappa_{min})=(0.04;0.01)$ simulated with LBM with $N_y=16$ nodes (crosses) and finite difference method with $512$ nodes (lines) at different applied pressure.
	Bottom: Dashed line represents the distribution of the local critical pressure and the horizontal continuous line is the different applied pressure. 
	As expected, the velocity is non-zero where the pressure gradient is higher than the local critical pressure.	
	\label{velocity_profile_error}}
\end{figure}

\begin{figure}
	\includegraphics[width=0.9\linewidth]{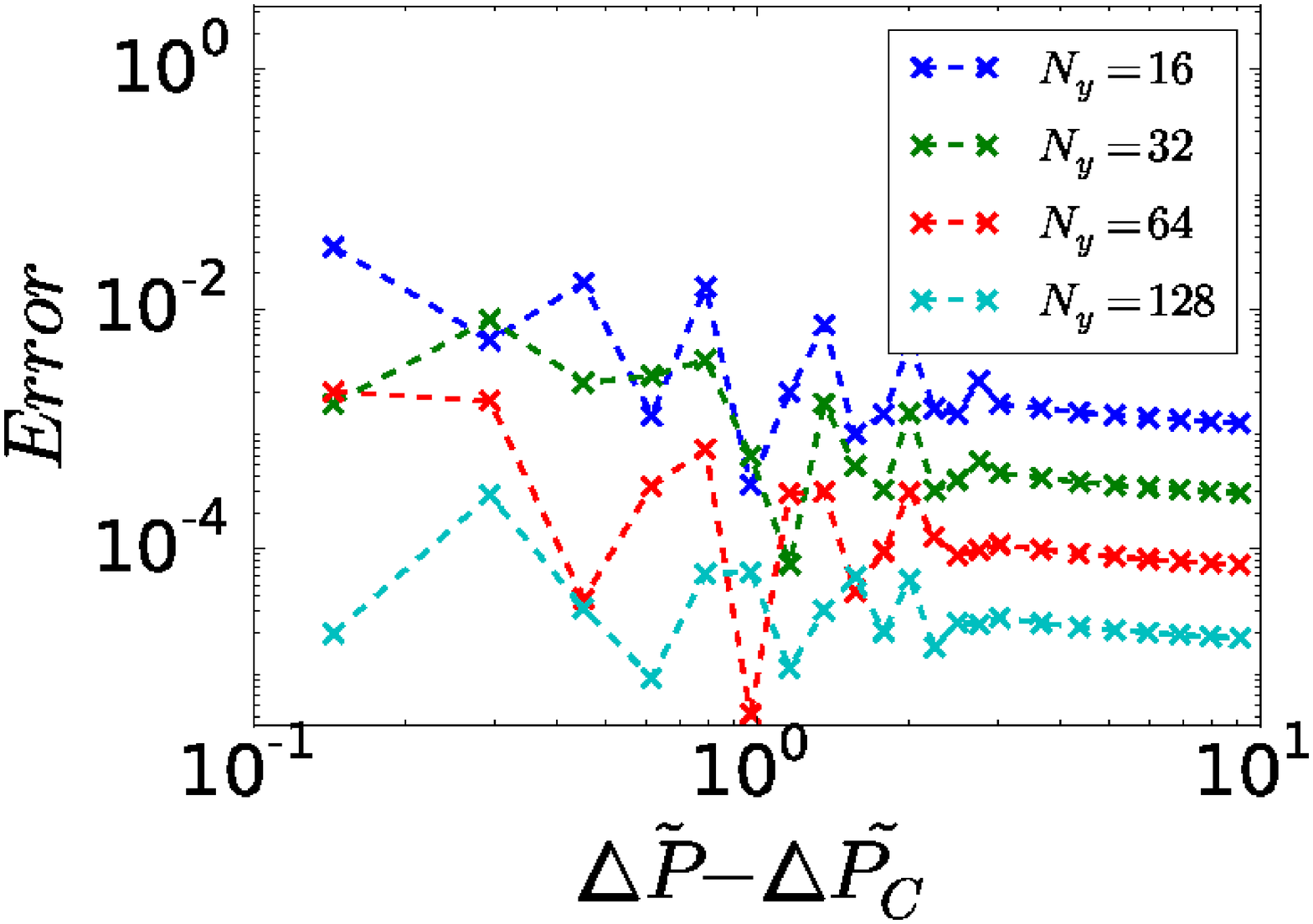}
	\caption{Relative total flow rate error for different discretisation number nodes $N_y$ as function of the applied pressure drop.\label{flow_rate_error}}
\end{figure}

In Fig. \ref{velocity_profile_error}, we have plotted the velocity profile for different applied pressure field for a fixed resolution $N_y=16$.
As we  can see in this figure, the difference with the finite difference is quite satisfactory.
We also compute in Fig. \ref{flow_rate_error}  the error difference between the total flow rate $Q_{LBM}$ and $Q_{FD}$ from the Lattice Boltzmann and the finite difference method respectively: $$Err=\frac{Q_{LBM}- Q_{FD}}{Q_{FD}}.$$
The error remains quite low. The maximum of errors is reached close to the critical pressure where both solutions tend to zero.

Considering eq. \eqref{eq:correlation} with $\lambda=5$, we can estimate  that our simulations correspond to a periodicity of $N_y \simeq 20$. We expect thus an error of $1-2\%$ close to the first path opening, and of less than $0.1\%$ above.

\section{Results}
\subsection{Problem description}

We simulate the flow in domain size of  $4096\times 1024$. The permeability distribution is log-normal with a mean $k_0=e^{-1}$, a correlation length $\corrsize = 5$ and a mean square deviation $\std^2$ ranging from $0.1$ to $3$. The constant $A$ was set to $A=10^{-5}$.
A constant pressure difference $\Delta P$  is then imposed at the inlet and outlet of the domain.
After the simulation has converged, we  compute the mean flow rate: $\tilde q = \frac{1}{V}\int u dV $, where $V$ is the domain size.

\begin{figure}
	\includegraphics[width=0.23\linewidth,trim={2cm 0 2cm 0},clip]{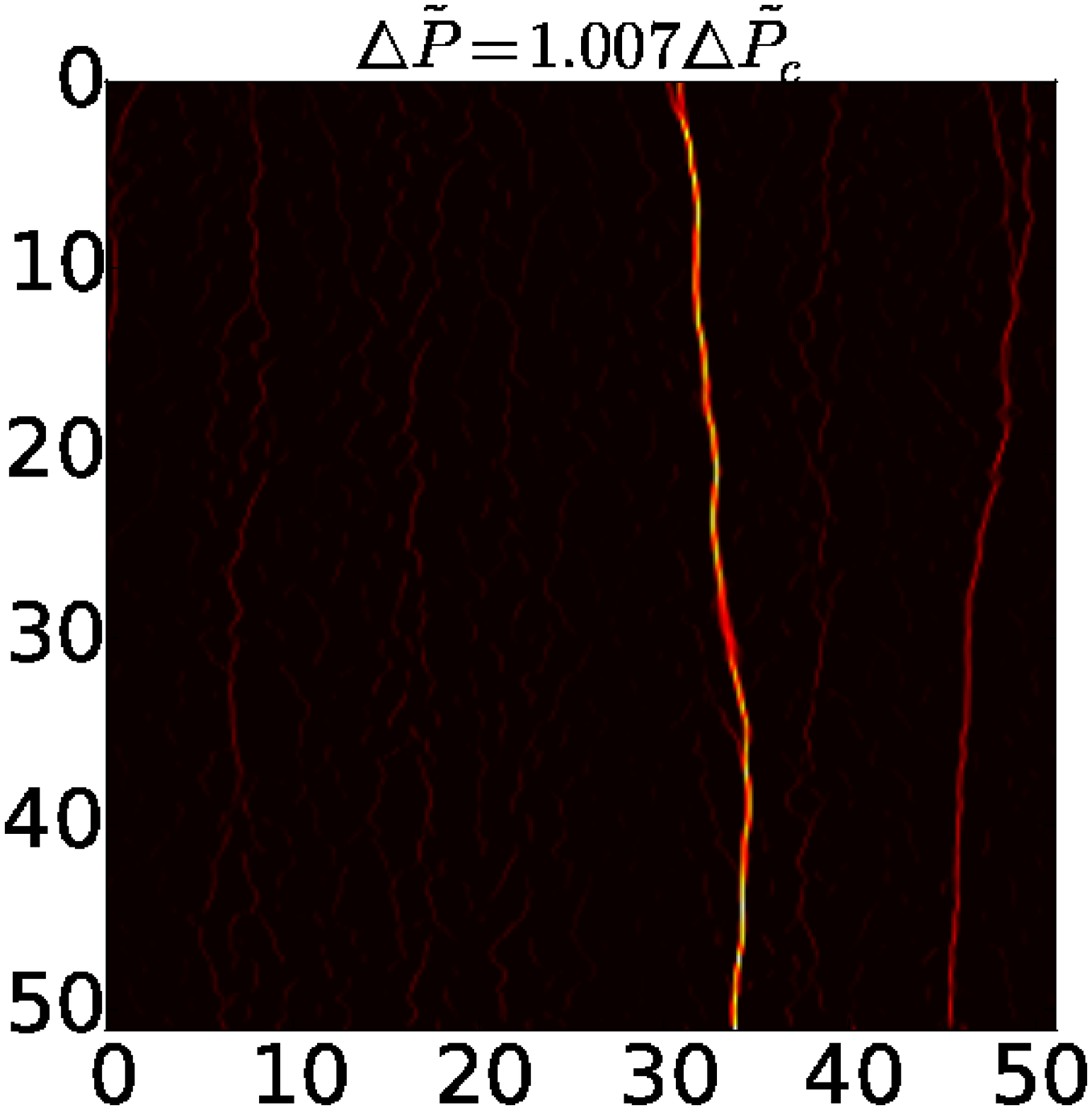}
	\includegraphics[width=0.23\linewidth,trim={2cm 0 2cm 0},clip]{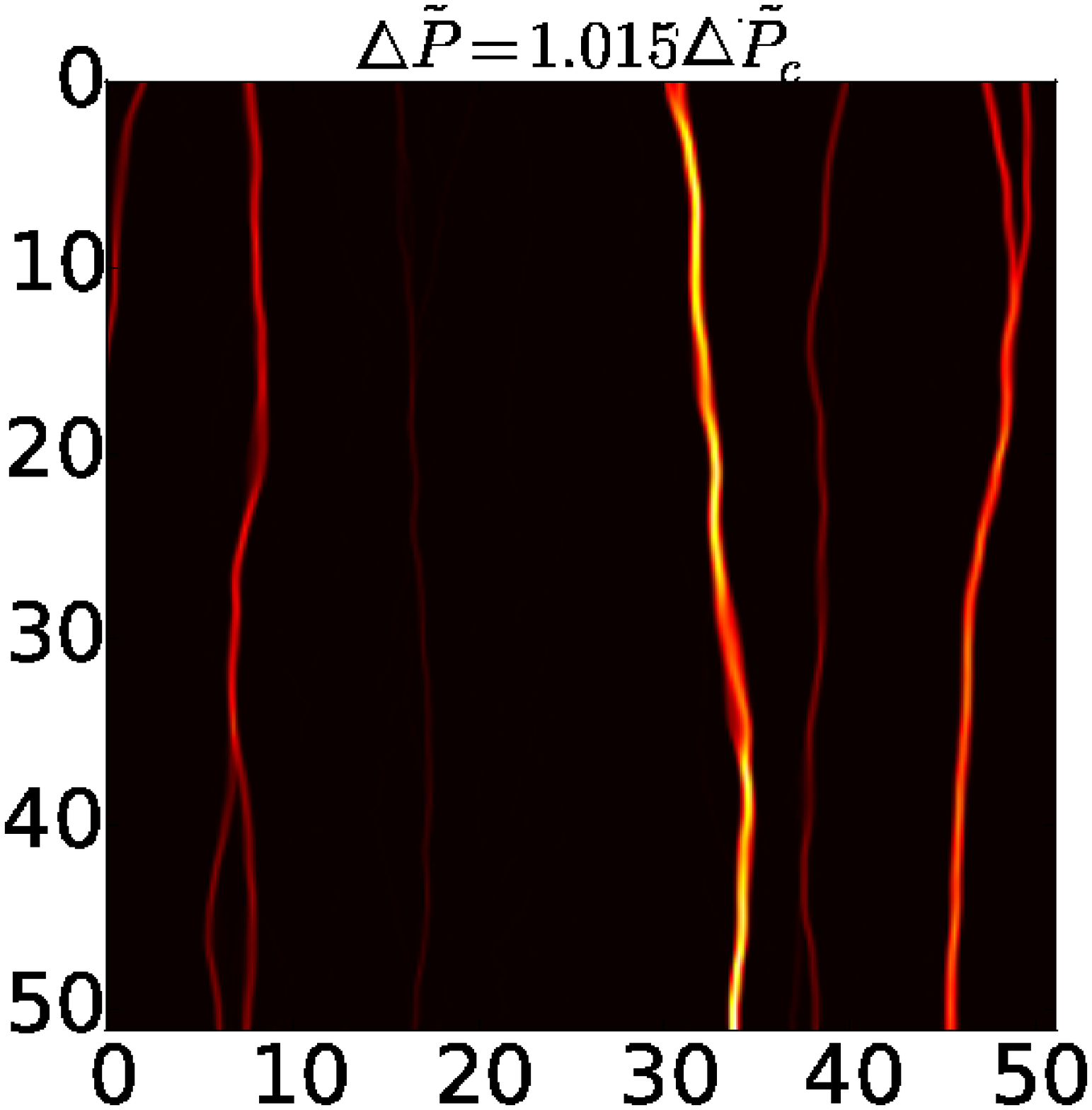}
	\includegraphics[width=0.23\linewidth,trim={2cm 0 2cm 0},clip]{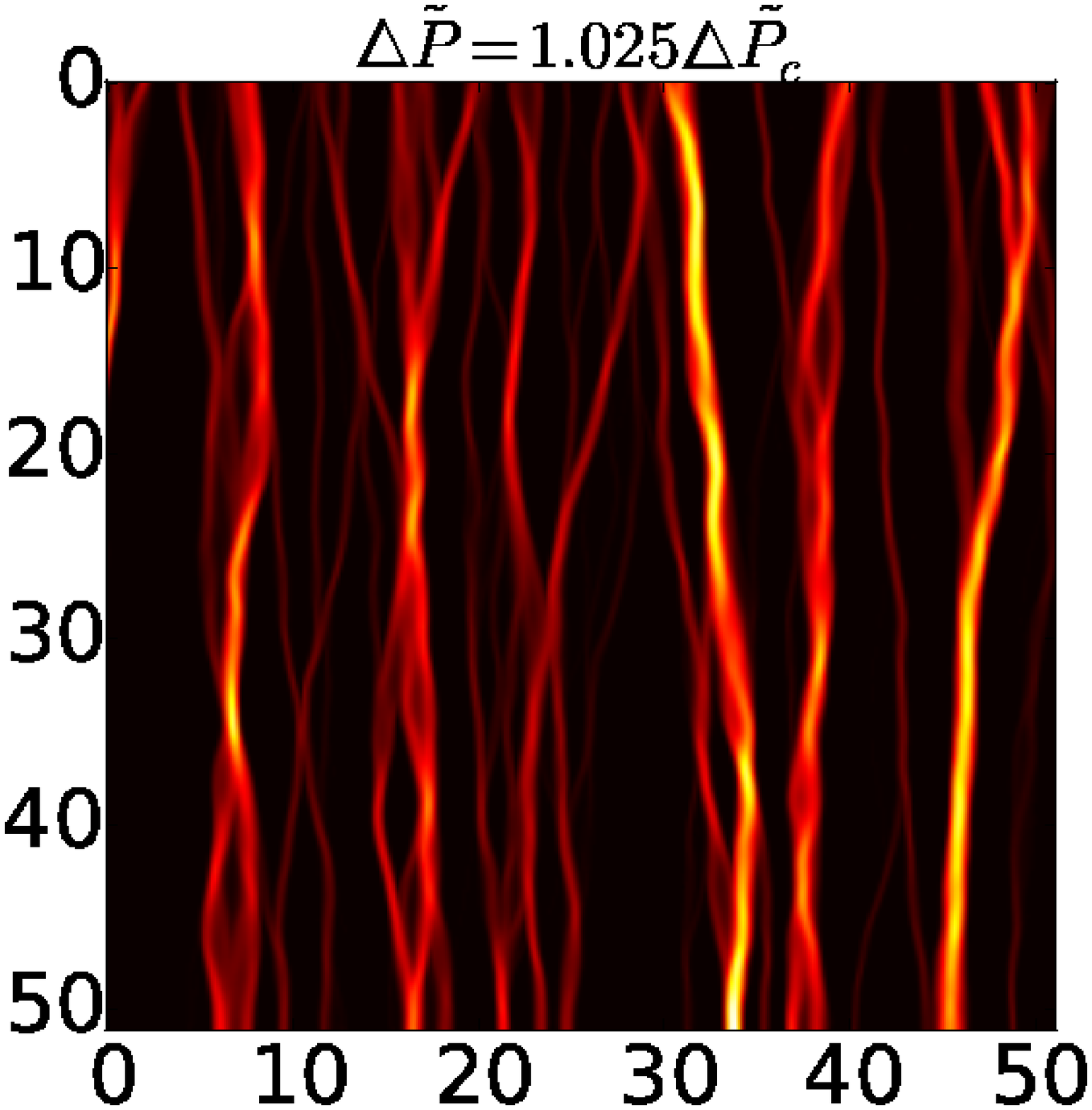}
	\includegraphics[width=0.23\linewidth,trim={2cm 0 2cm 0},clip]{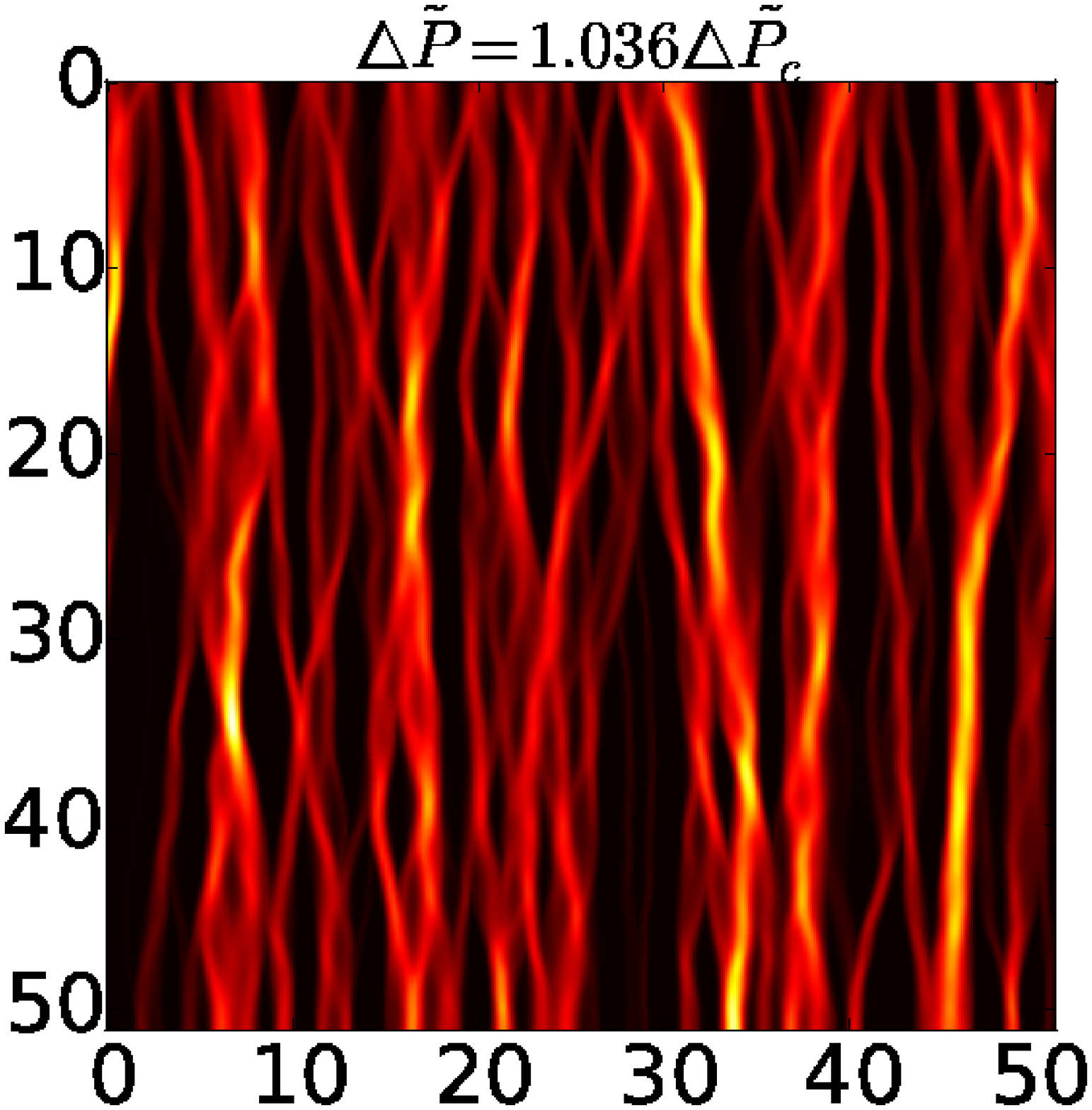}
	\includegraphics[width=0.23\linewidth,trim={2cm 0 2cm 0},clip]{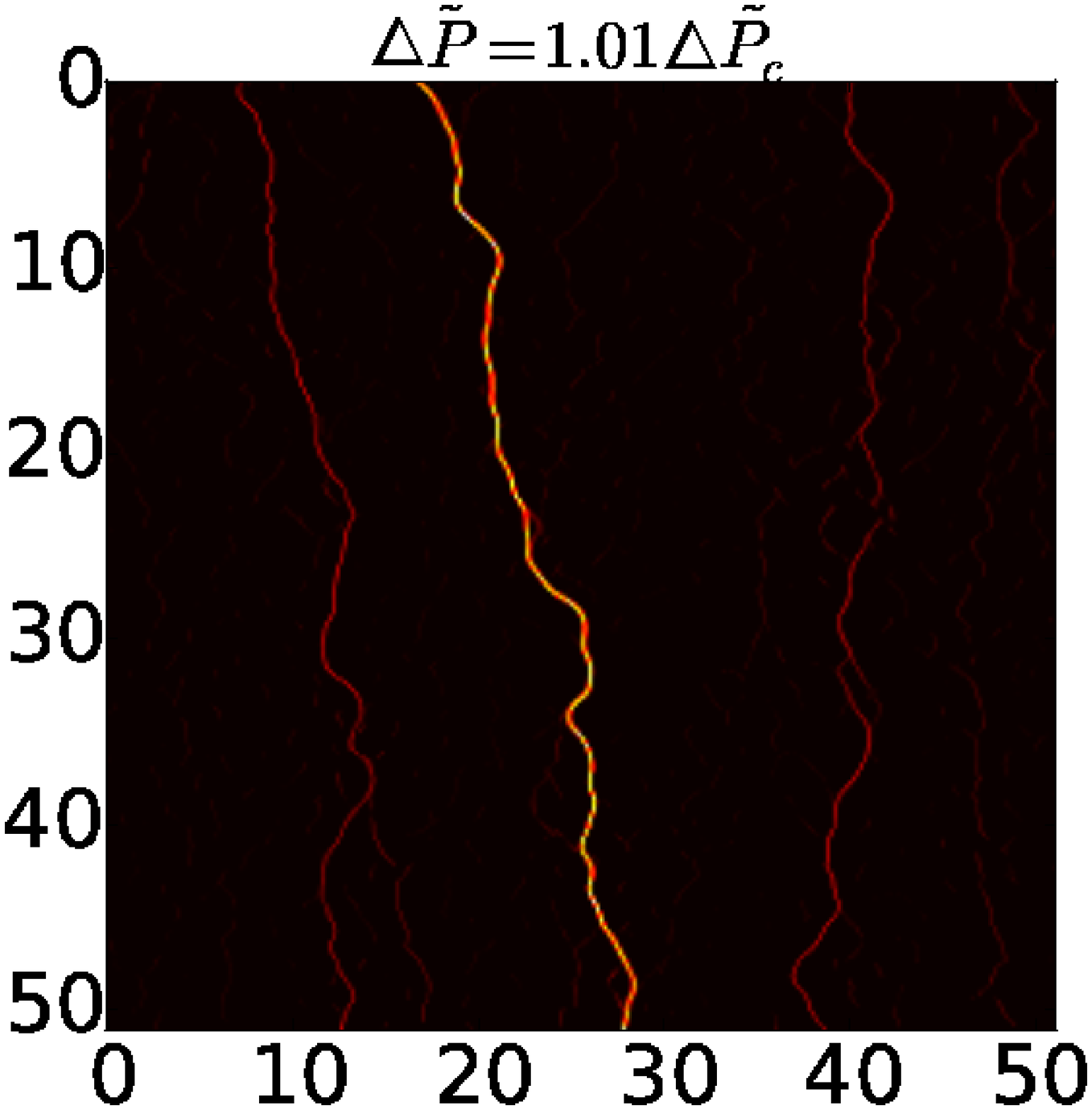}
	\includegraphics[width=0.23\linewidth,trim={2cm 0 2cm 0},clip]{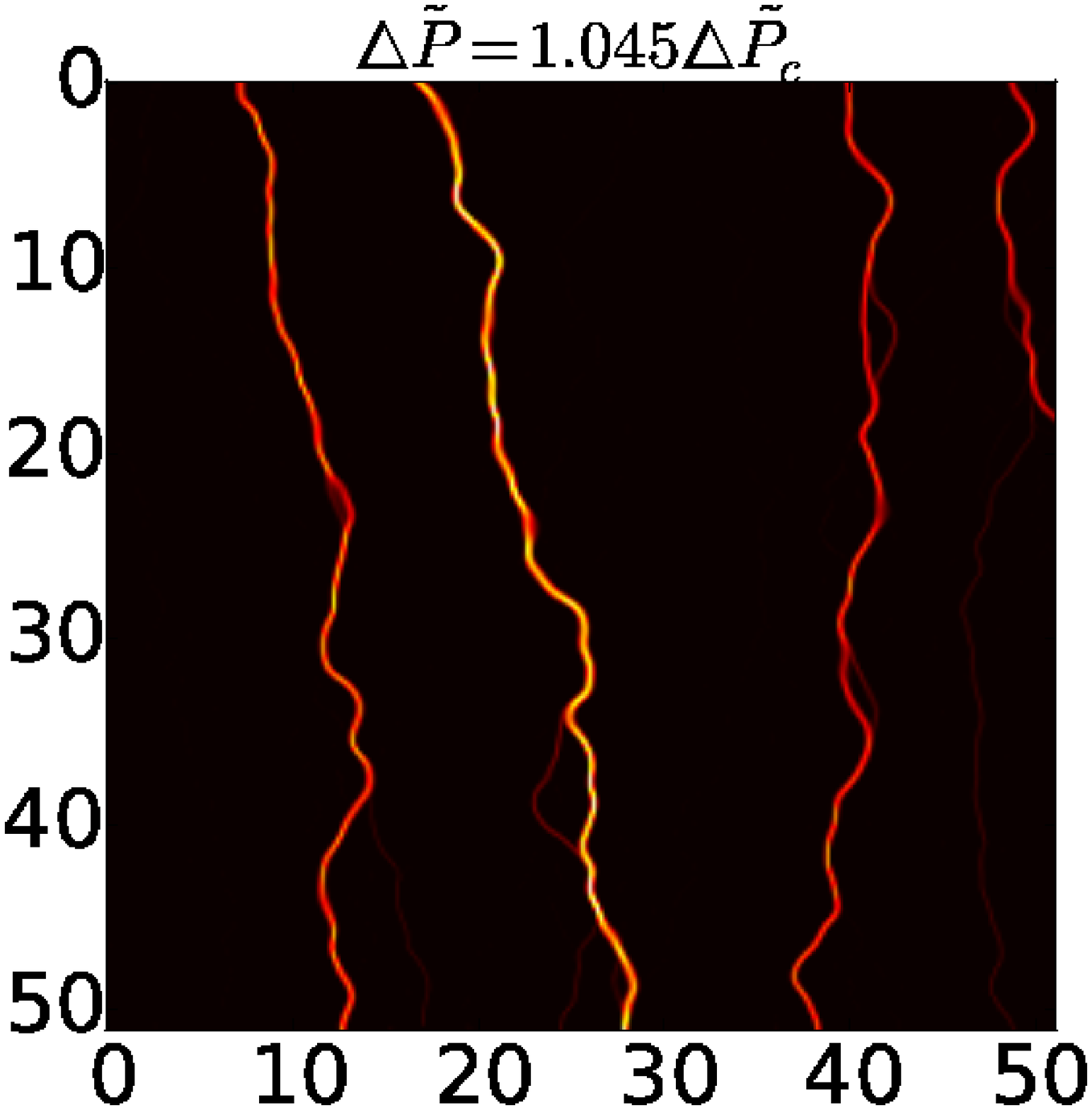}
	\includegraphics[width=0.23\linewidth,trim={2cm 0 2cm 0},clip]{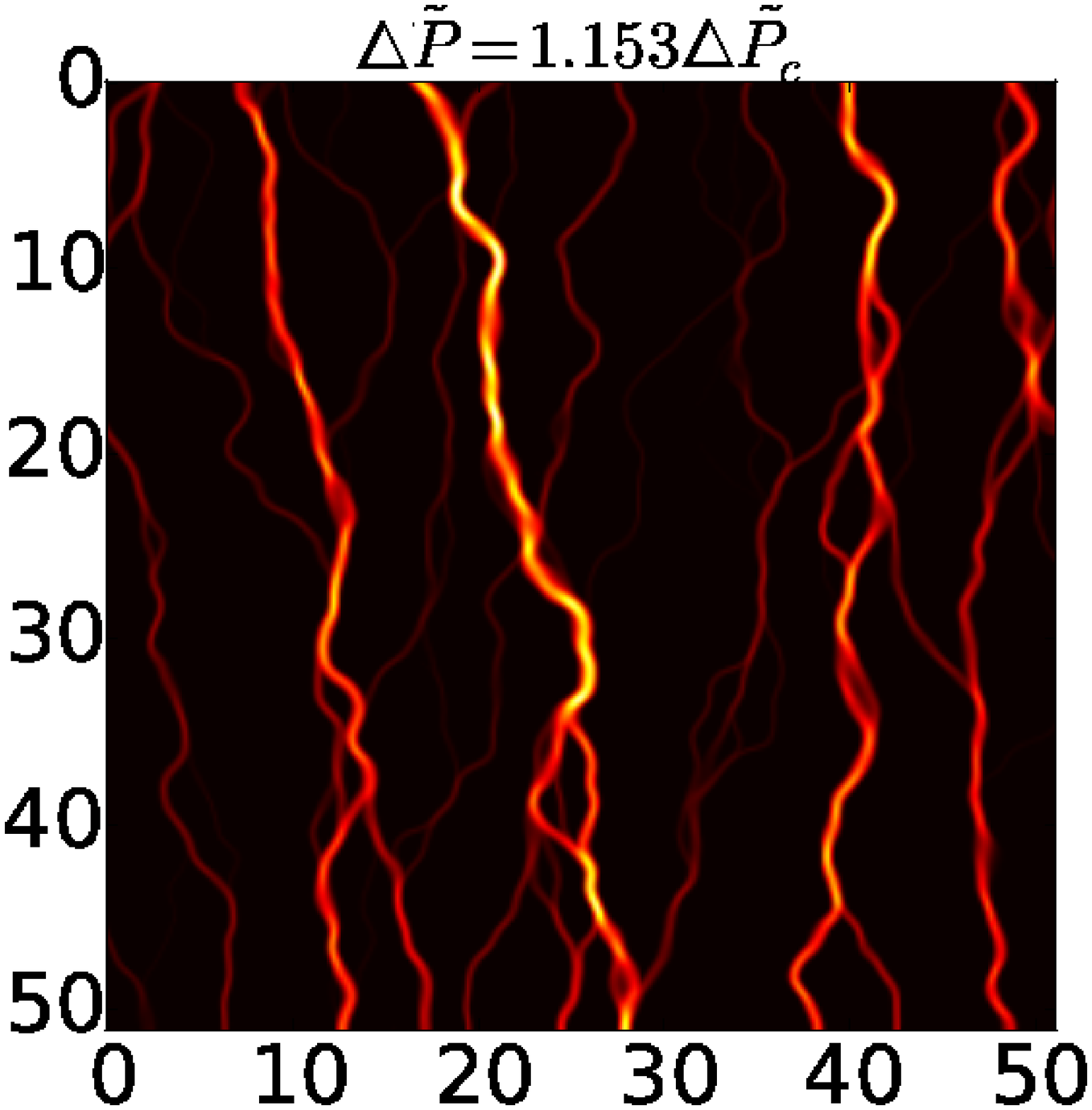}
	\includegraphics[width=0.23\linewidth,trim={2cm 0 2cm 0},clip]{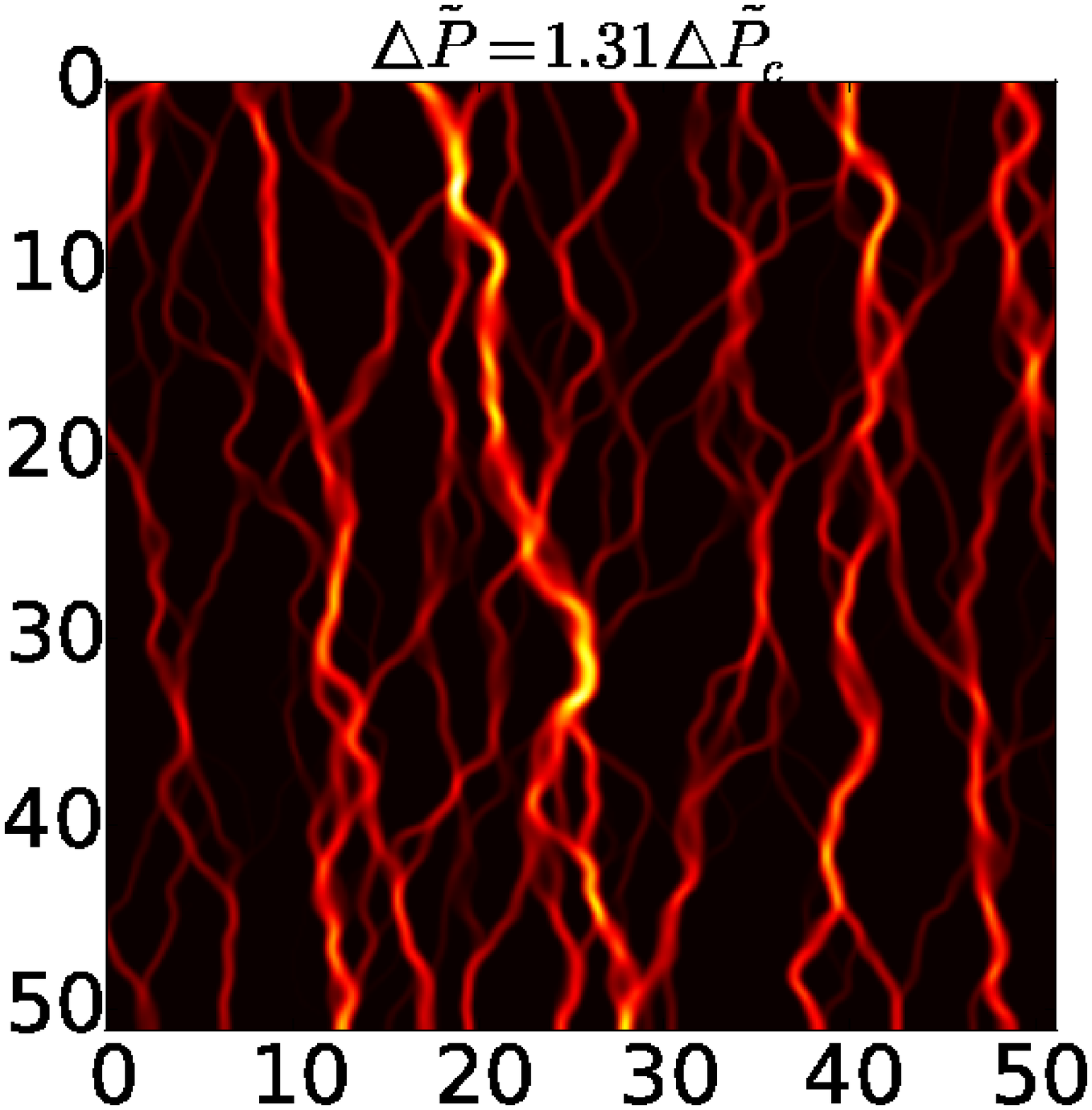}
	\includegraphics[width=0.24\linewidth,trim={1.8cm 0 2cm 0},clip]{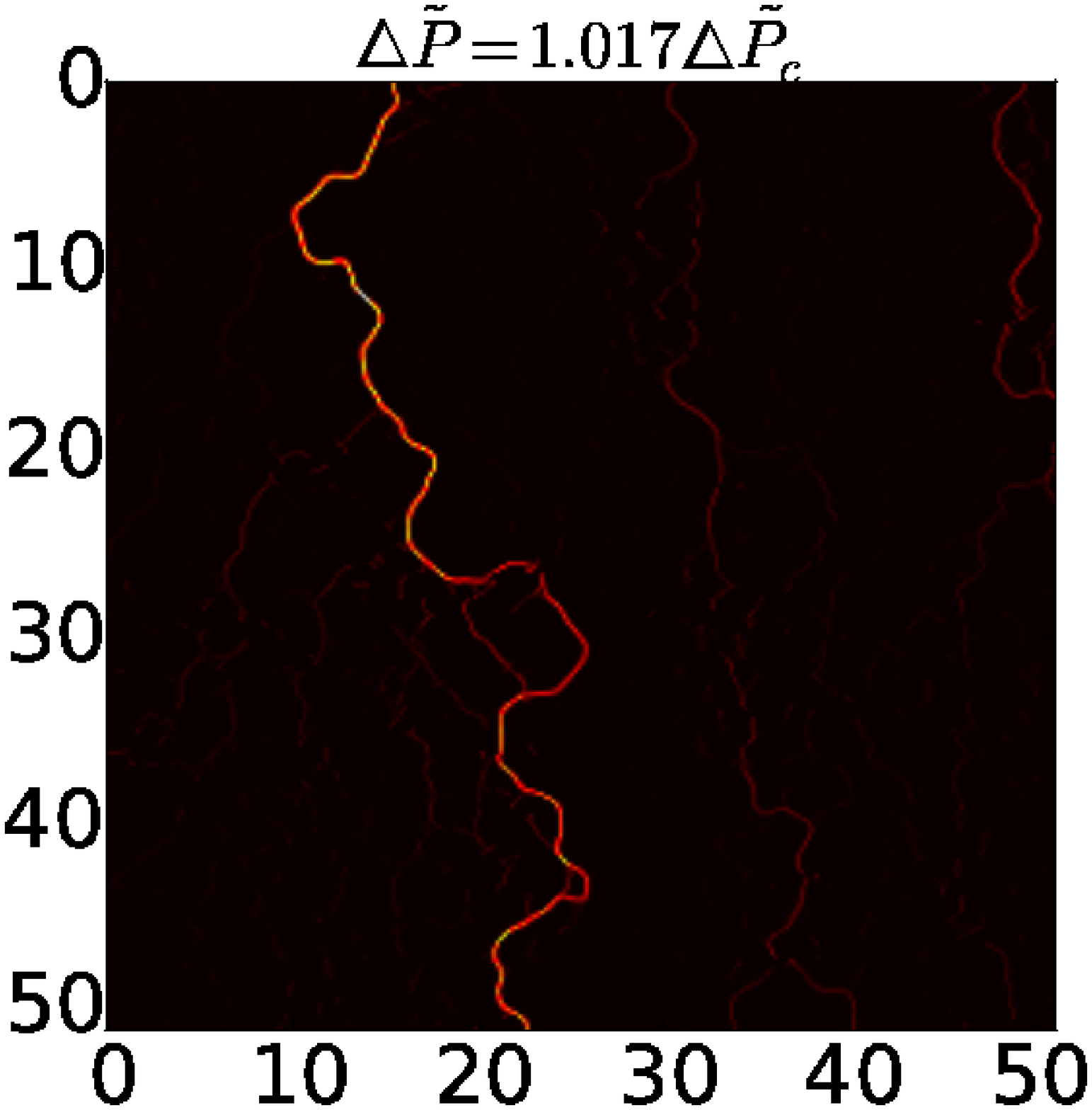}
	\includegraphics[width=0.24\linewidth,trim={1.8cm 0 2cm 0},clip]{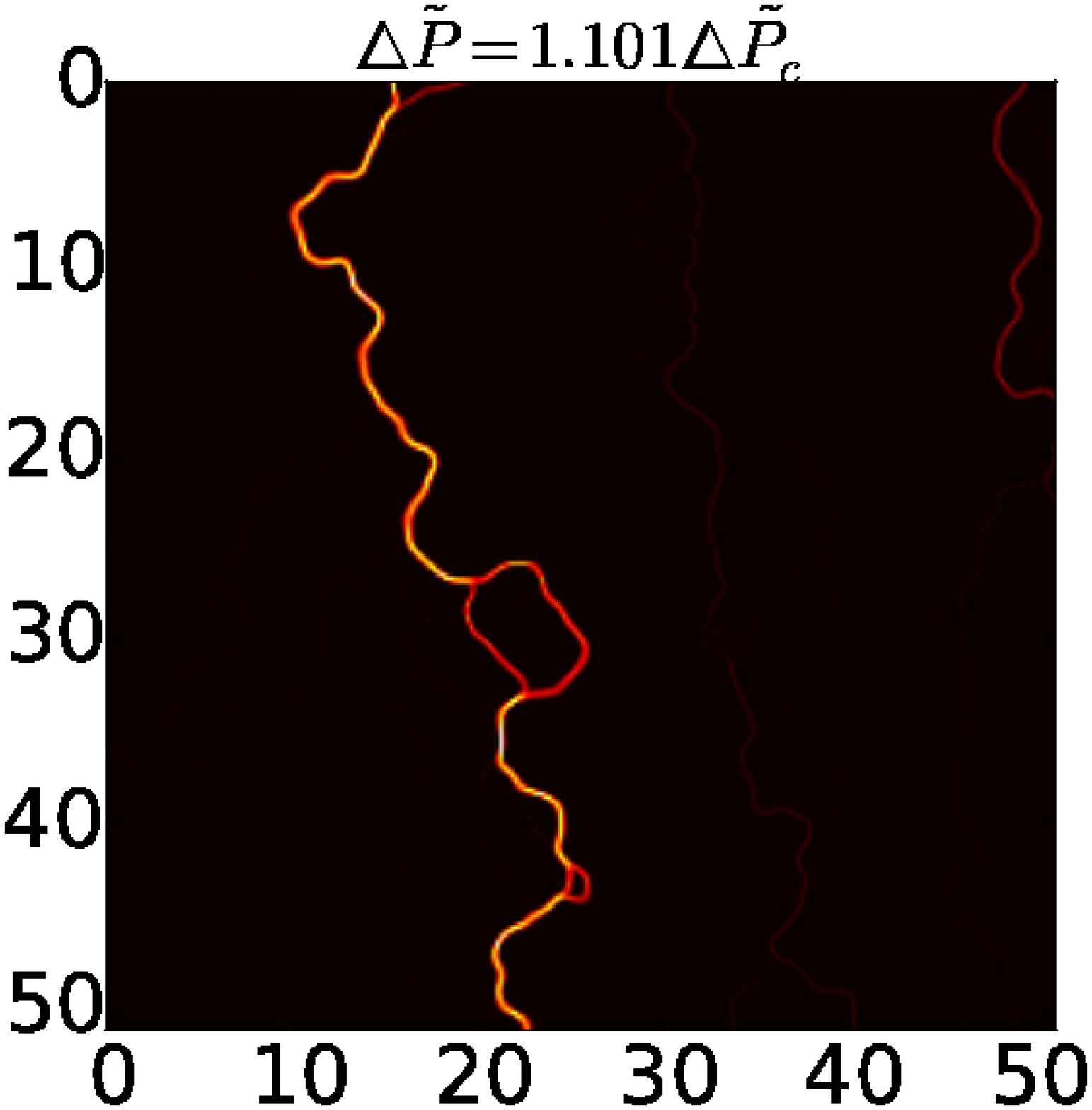}
	\includegraphics[width=0.24\linewidth,trim={1.8cm 0 2cm 0},clip]{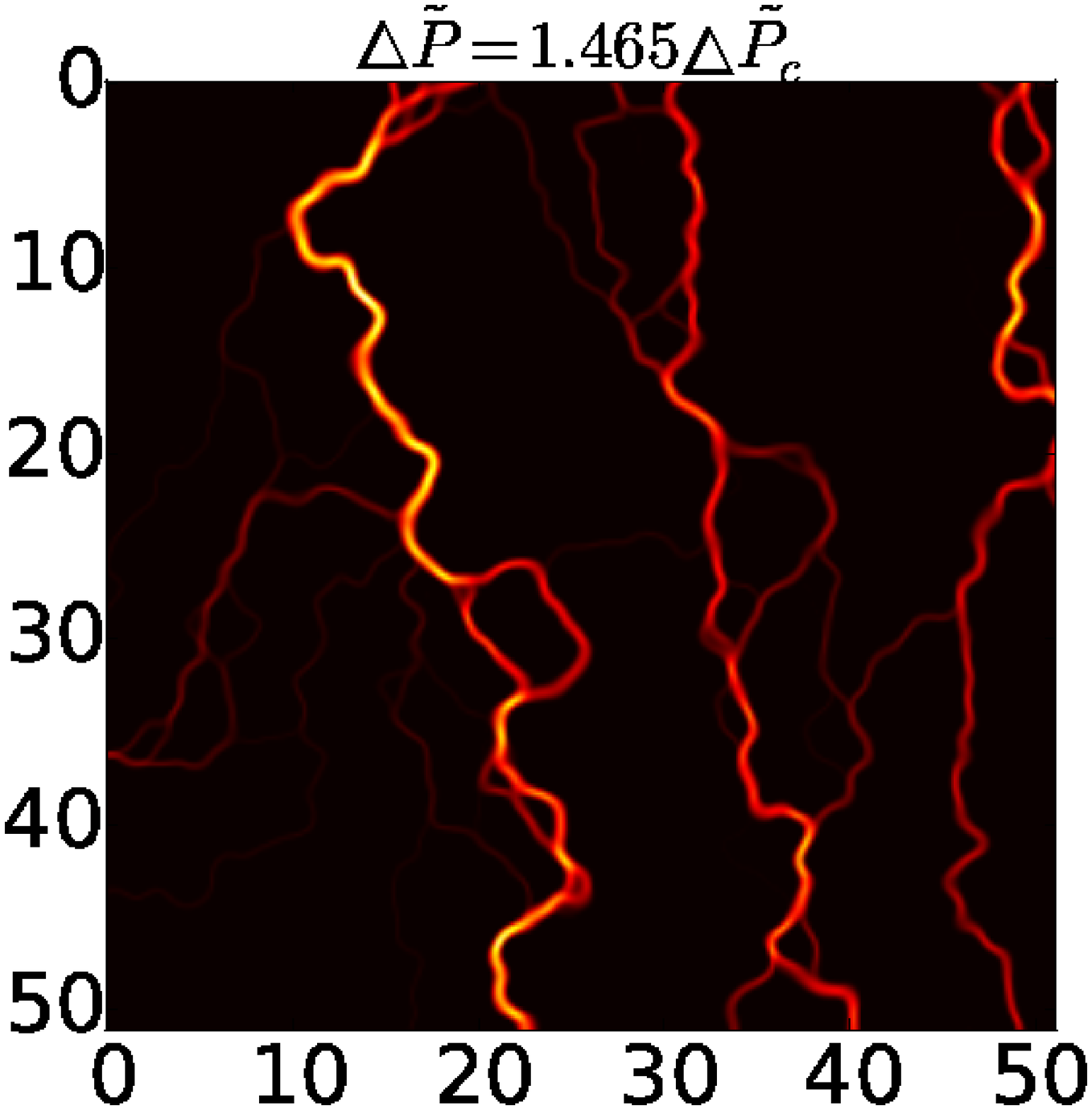}
	\includegraphics[width=0.24\linewidth,trim={1.8cm 0 2cm 0},clip]{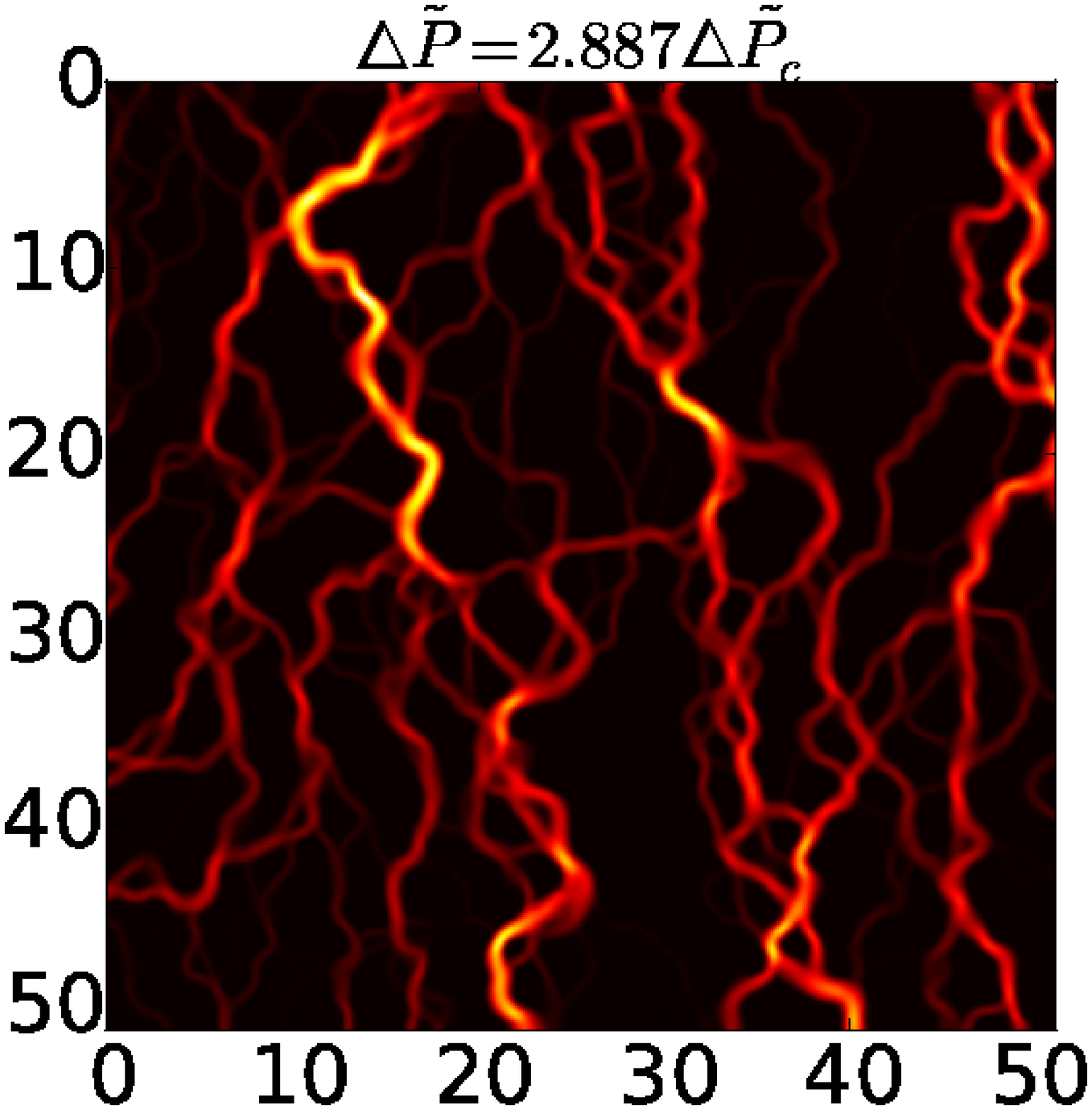}
	\caption{Snapshot of the velocity map at increasing pressures beyond $\aPcsyst$. Flowing paths are grey and white (color online), clusters are in black. 
	Different line correspond to different $\std$:	from top to bottom $\sigma^2=0.1$, 	$\sigma^2=1$, $\sigma^2=3$.
	The different columns are for  increasing applied pressure difference from left to right.
	\label{Velocity_map}}
\end{figure}

When the imposed pressure is too low, the whole domain is in the solid state which results in a zero mean velocity (very small in fact due to the capping of the viscosity).
Above this macroscopic threshold, the fluid then starts to flow in few channels.
Figure \ref{Velocity_map} displays  snapshots of the flow field  for different $\std$ and different  applied pressure above this threshold.
As it can be seen, close to this threshold the number of flowing paths is small.
In principle, it should reduce to a single path but, due to numerical precision, the precise critical point is difficult to determine, particularly for low $\std$.
As the pressure is increased, the number of flowing paths increase with the applied pressure.
We also observe in these snapshots that the flowing paths seem to be more tortuous as the heterogeneity parameter, $\sigma$,  is increased.

In the following, we will first present the  evolution of the mean flow characteristics  by increasing the pressure. Then we will study some geometrical aspect of the flow field.

\subsection{Onset of flow}
The first characteristic is the critical global pressure difference required to initiate the flow.
Because of the yield stress property, we observe a minimal pressure drop $\aPcsyst$ below which there is no flow (within the regularization error bar).
As already described in other papers \cite{roux87,talon13a, hewitt16}, this pressure corresponds to the appearance of the first flowing path.
It can be formally defined as being the path which has the minimal critical pressure:
\begin{equation}
\label{eq:critical_path}
\aPcsyst = \min_{\cal C} \int \aGP(\vec r) ds
,
\end{equation}
 where the path $\cal C$ are taken among all possible paths connecting the inlet to the outlet.

Several comments must be made on this critical path. We expect that the path and the associated critical pressure should depend  on the disorder since they result from the competition of two opposite effects.

On the one hand, the minimal path tends to connect regions of low $\aGP(\vec r)$ (\emph{i.e.} high permeability).
But on the other hand, as indicated by eq. \eqref{eq:critical_path}, the total length of the path has a contribution.
The optimal path results thus from a balance between the search of the most permeable areas and the cost of increasing the length of the path.
 
This competion can clearly be observed in Fig. \ref{Velocity_map}, where we have plot the first path for different $\sigma$. At lower $\sigma$, the path is almost straight because the lower $\aGP(\vec r)$ regions are not low enough to compensate for an increase of length.
As $\sigma$ increases, the optimal path becomes more tortuous because the value of the lowest $\aGP(\vec r)$ regions decreases, which could be  worth the detour.

We may note that the tortuosity of the first path may not necessarily increase with the amplitude of the  disorder. A simple example would be to multiply the $\aGP$ field by a constant value (or by changing $A$),  which has the effect of increasing the standard deviation but not the shape of the optimal path (see eq. \eqref{eq:critical_path}).

Finally, we should also note that the path selection of eq. \eqref{eq:critical_path} is very close to a standart problem in statistical physics named directed polymer problem \cite{kardar87}, which consist in the finding
of a minimal path in an energy lanscape.
The minimization is however perform among "directed" paths, meaning that their slope is bounded to avoid any overhangs.
In this context,  the directed polymer is known to be self-affine with a roughness exponent equal to $2/3$.
This exponent is "universal" in the sense that it does not depend on the distribution of the energy (see \cite{kardar87, halpin-healy95}), provided that the distribution is not too "extreme" (like power law, fat tail, distributions \citep{gueudre15}).
The roughness of our flowing paths will be discussed in section \ref{sec:geometrical}.

\begin{figure}
	\includegraphics[width=0.9\linewidth]{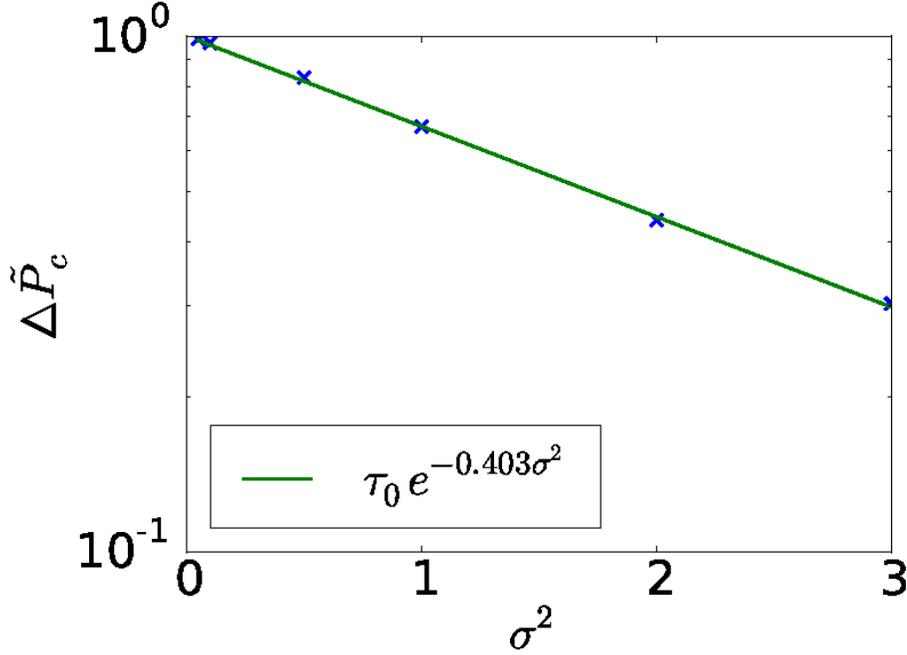}
	\caption{Semilog plot of th global critical pressure as function of the heterogeneity parameter $\std$. As we can see, this pressure exponentially decreases with $\aPcsyst(\std=0)=1$.
	 \label{fig:Critical_pressure_sigma}
	 }	\end{figure}

In Fig. \ref{fig:Critical_pressure_sigma}, we plot the value of $\aPcsyst$ as function of the amplitude of the disorder.
As expected, $\aPcsyst$ decreases with the amplitude of the disorder, starting from $\aPcsyst(\sigma=0) =  \aGP \alength$, where $\alength$ is the dimensionless total length of the system.
Moreover, as indicated in the figure, the critial pressure seems to follow a gaussian function with $\std$.
 
Note that for low value of $\sigma$, the critical pressure and the path associated to it become more difficult to determine precisely.
Indeed, since the disorder is small, all the possible paths are quite similar in term of pressure threshold.
As a result, all flow paths appear very quickly in a narrow pressure range.

\subsection{Flow regimes}

\begin{figure}
	\includegraphics[width=0.9\linewidth]{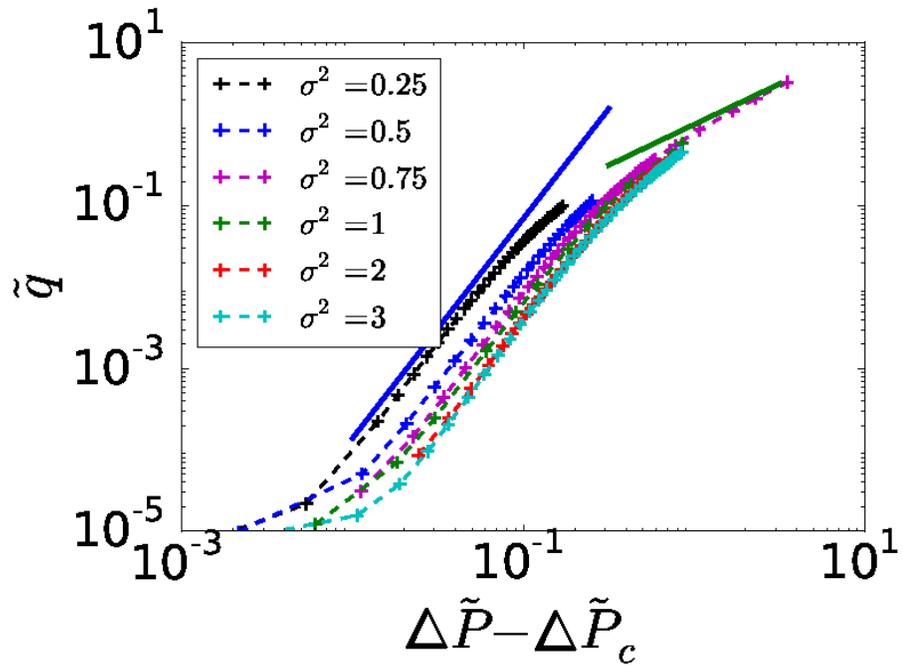}
	\caption{Flow rate as function of the pressure difference $\tilde{q}(\Delta \tilde{P})$ in a log-log scale for different $\std$.
The intermediate continous line corresponds to a power-law of exponent $2.8$, the last one corresponds to a linear law.
	\label{flow_rate}}
\end{figure}

We now study the evolution of the flow rate by increasing the pressure difference above the threshold.
We have plotted in Fig. \ref{flow_rate}, the mean flow rate $q$ as function of the applied pressure substracted by the critical pressure $\Delta \tilde{P} - \aPcsyst$.
As it can be seen, the mean debit follows a power-law  over a certain range of pressure:
\begin{equation}
\tilde{q} = B (\Delta \tilde{P} - \aPcsyst)^\flowpower.
\end{equation}
More remarkable,  we note that the exponent seems to be constant with the heterogenity paramter $\std$ (within the error bar) $\flowpower = 2.8 \pm 0.05$.
 
\begin{figure}
	\includegraphics[width=0.9\linewidth]{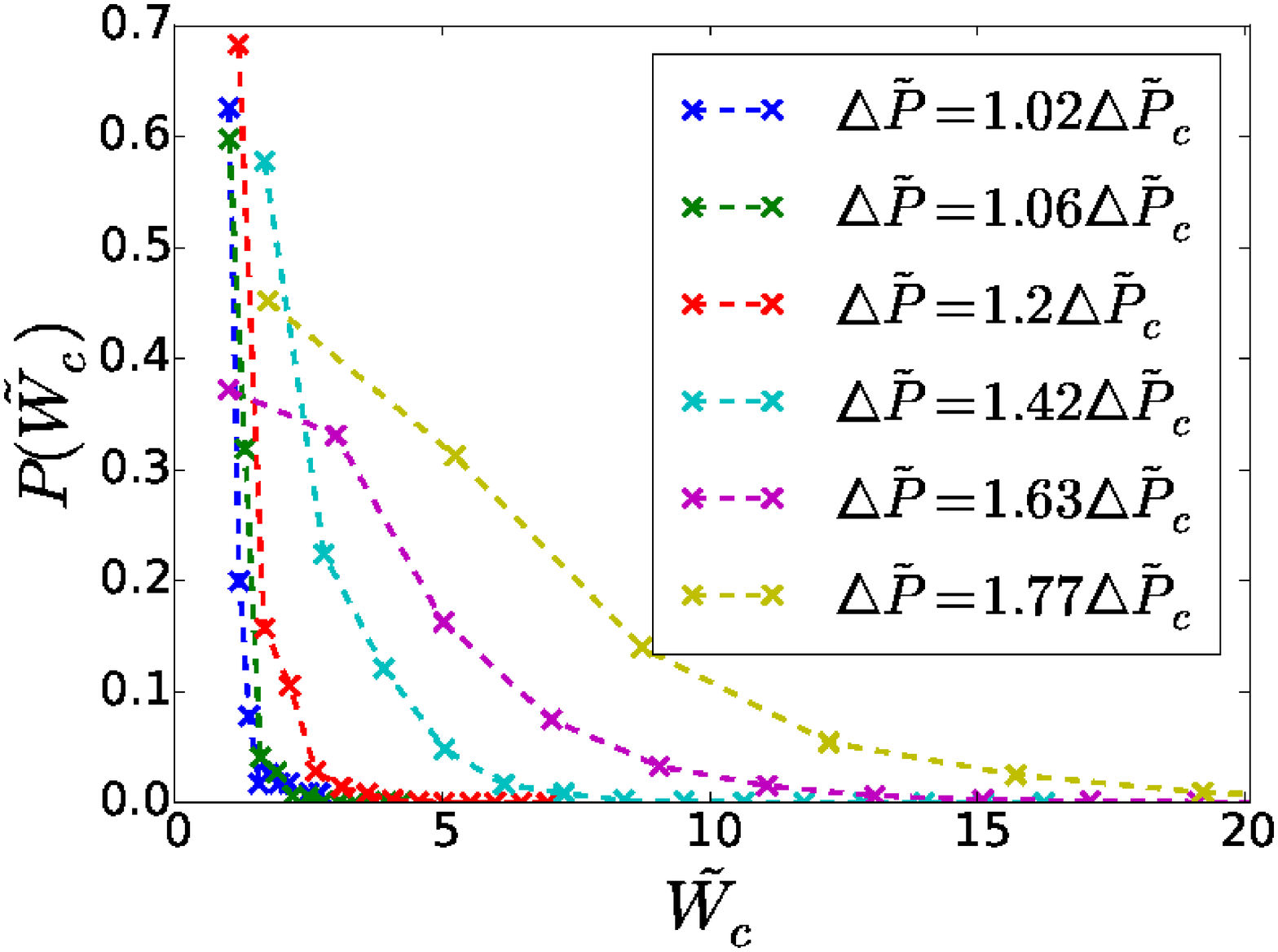}
	\caption{Probability distribution of the channel width,  ${W_c}$, for different applied pressure difference.	\label{channel_width}}
\end{figure}

This flow behaviour is reminiscent of what has been observed at the pore scale  \cite{chevalier15a}.
However, there is a major difference that lies in the value of the exponent which is significantly higher ($\flowpower \simeq $ 2 at the pore scale).
 We believe that this is due to the contribution of the increase in the number of flow paths but also to the increase of the flow rate in each of them.
 As in the case of pore scale, a significant contribution is due to the increase in the number of flow paths with the applied pressure.
However, while at the pore scale, the flow rate of each path is expected to increase linearly with pressure, this is not necessarily the case at the macroscopic level because the paths can also widen.
 Indeed, since the permeability field and the limit pressure are correlated, it is also expected that regions near open paths will also be easier to flow.
 The width of each individual path therefore increases with pressure, similarly to the stratified permeability distribution case  considered in the validation section.
This effect was quantified in Fig. \ref{channel_width} where we measured the distribution of the channel width $\tilde W_c$ as a function of the  applied pressure.
We can clearly see the broadening of the distribution with the increase in pressure.

Finally, for large enough pressure, once most of the domain is flowing, the number of the paths and their width cannot increase anymore, the flow rate then recover  a linear behavior $\tilde{q}\propto (\Delta \tilde{P} - \aPcsyst)$.

It should be noted that the amplitude of the disorder modifies the pressure range over which the intermediate regime is observed.
Indeed, the decrease in $\std$ results in a faster transition to the linear regime.
The effect is particularly strong for $\std = 0.1$, where most of the flow paths open very quickly after the first one. As we have discussed above, since the environment is more homogeneous, most of the possible paths are roughly equivalent.

\subsection{Geometrical propeties of the flow field \label{sec:geometrical}}

We now study the statistical properties of the flow field. If the main quantity of interest should be the flow paths, their characteristics (branch length, rugosity, etc.) are quite difficult to define and measure precisely.
Hence, it is more convenient to focus on the fluid at rest. We  study the statistics of the non-flowing fluid clusters, i.e. the areas of fluid at rest that are surrounded  by flowing channels.
Using a standard Hoshen-Kopelman algorithm on the velocity map allows us to determine the cluster of fluid at rest. For each cluster, we then extract its size $\tilde{S}$.
Its length $\alength$ and width $\awidth$ are then determined by fitting it into a rectangle. Where the length, $\alength$, is defined as the dimension along the flow direction and the width, $\awidth$, as the dimension transverse to it.

\subsubsection{Probability distribution}

\begin{figure*}
	\includegraphics[width=0.49\linewidth]{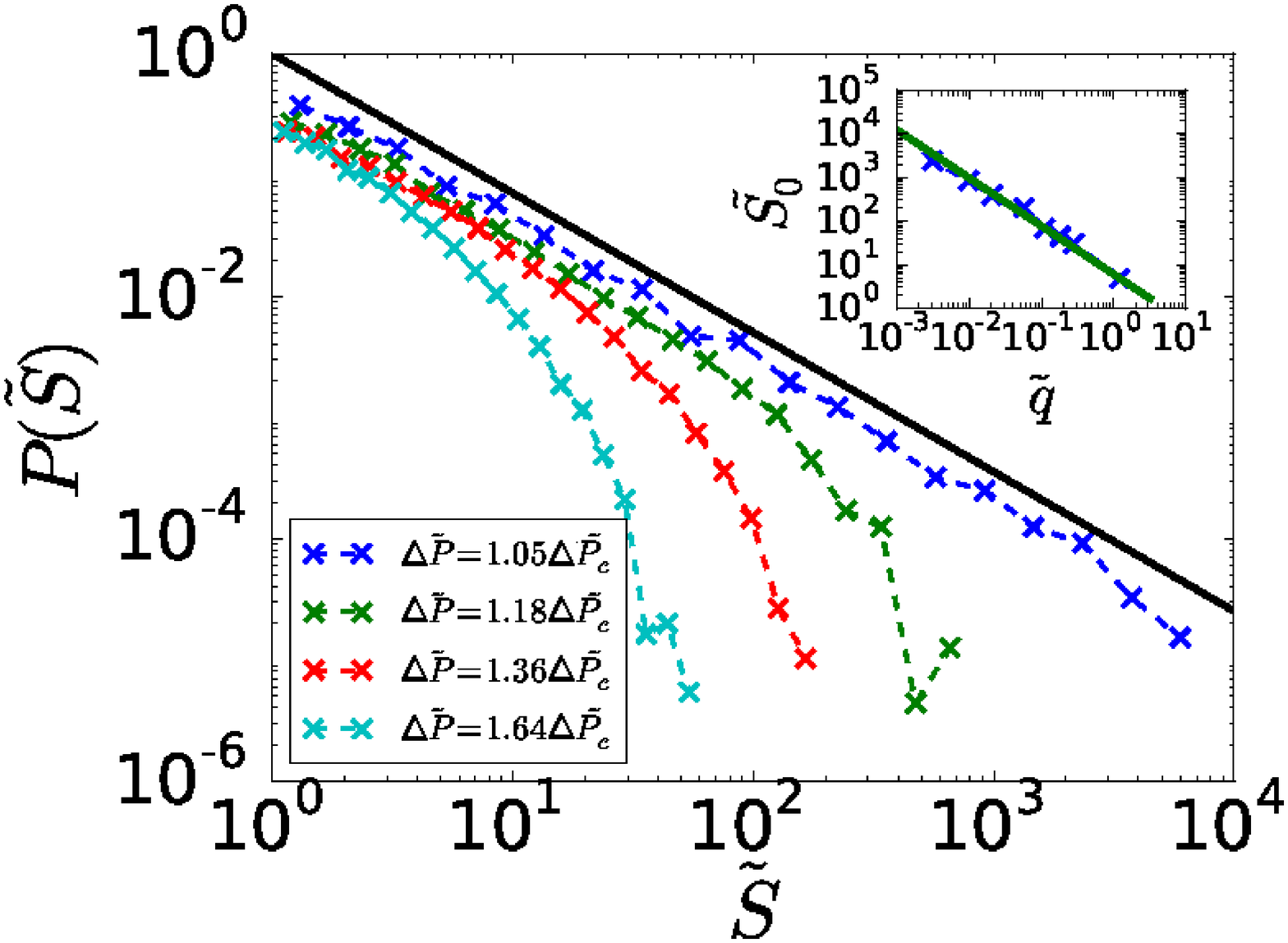}
	\includegraphics[width=0.49\linewidth]{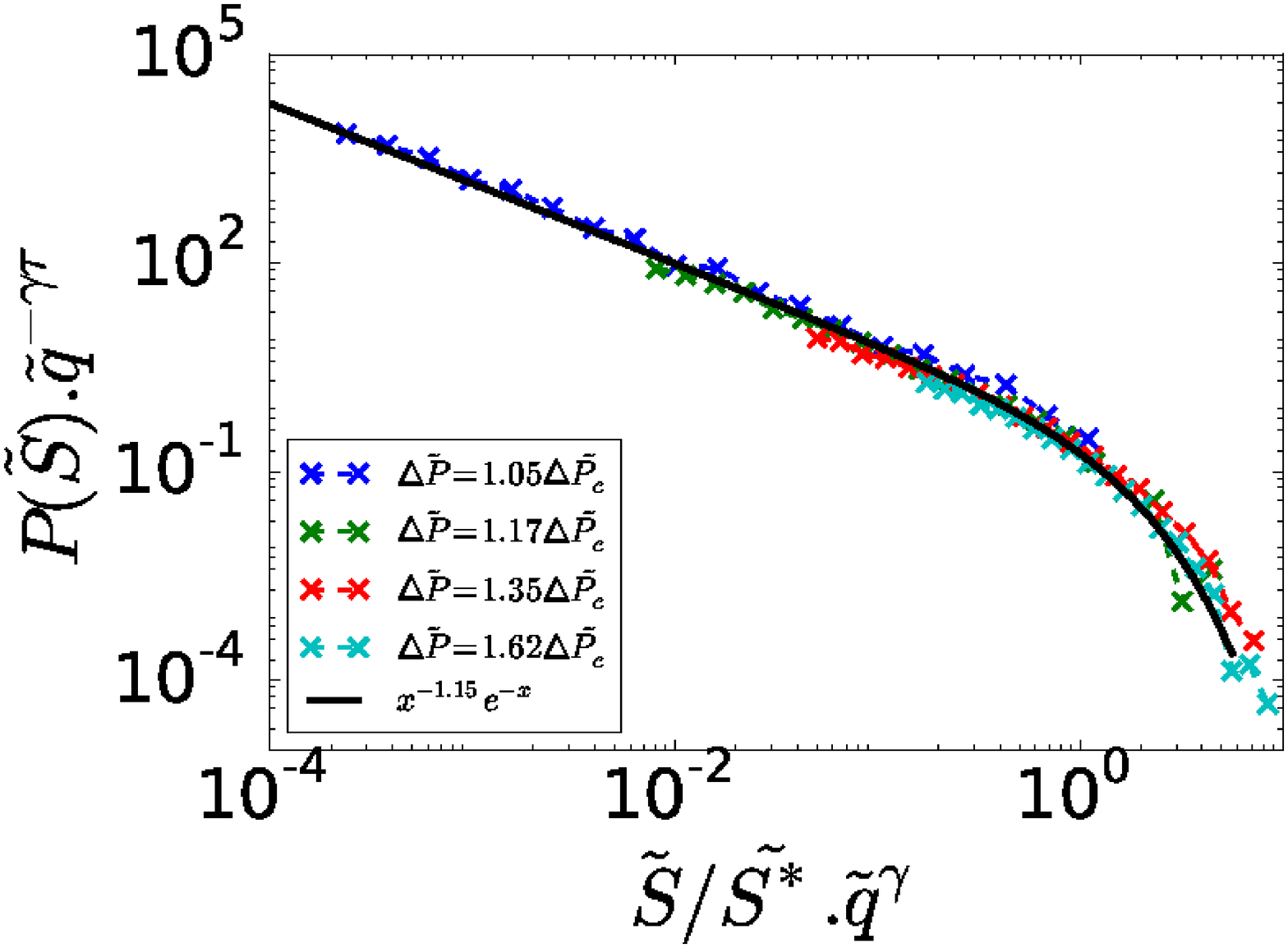}
	\caption{Left: Cluster size probability distribution versus cluster size $P(\tilde{S})$ and(inset) clusters cut size versus flow rate $\aScut(\tilde{q})$.
		Right: Normalized and collapsed cluster size probability distributions: $P(\tilde{S}).\tilde{q}^{-\Spower \Scutpower}$ verus $\tilde{S}.\tilde{q}^{\Scutpower}$.\label{P_S}}
\end{figure*}

\begin{figure*}
	\includegraphics[width=0.3\linewidth]{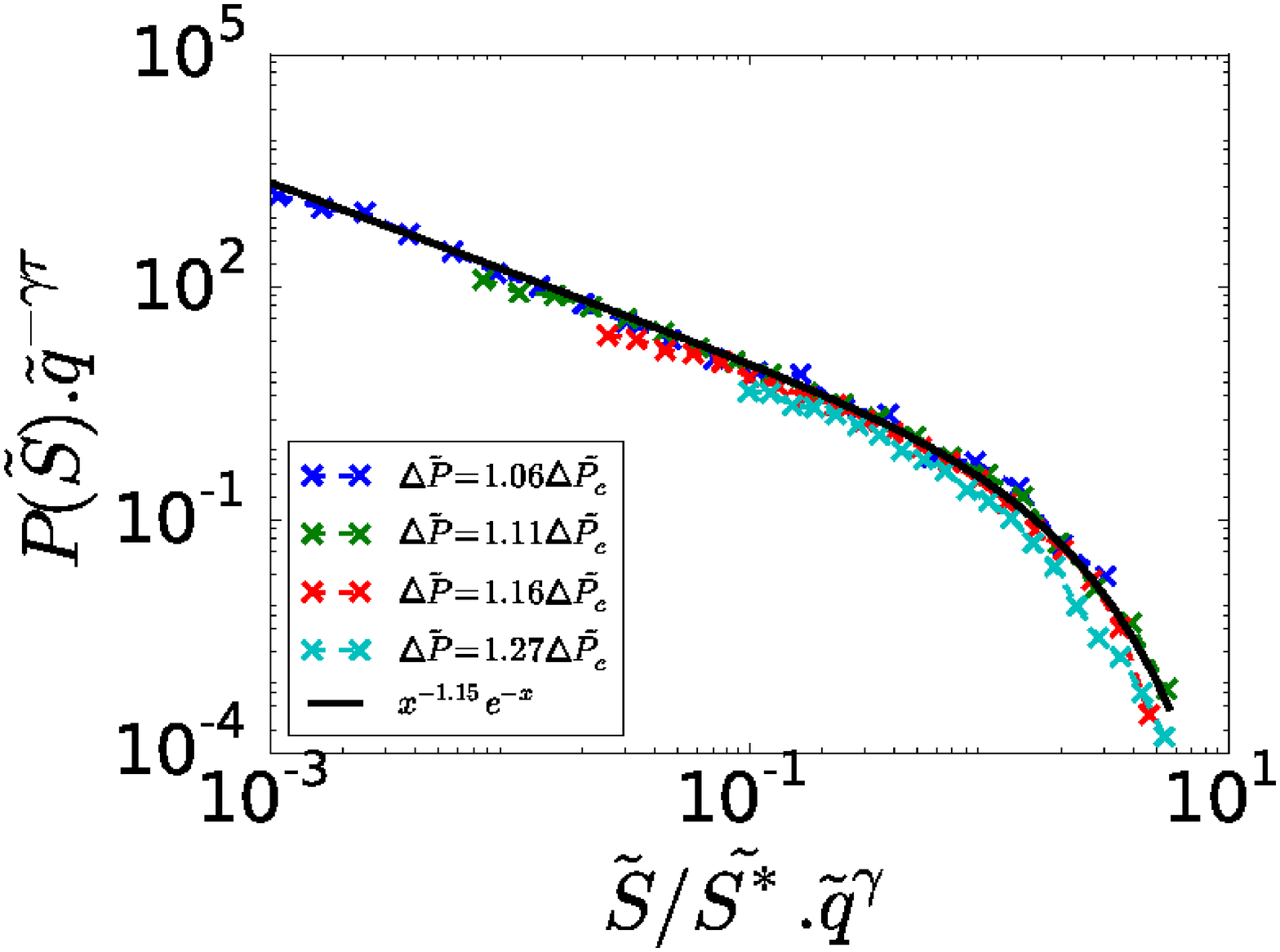}
	\includegraphics[width=0.3\linewidth]{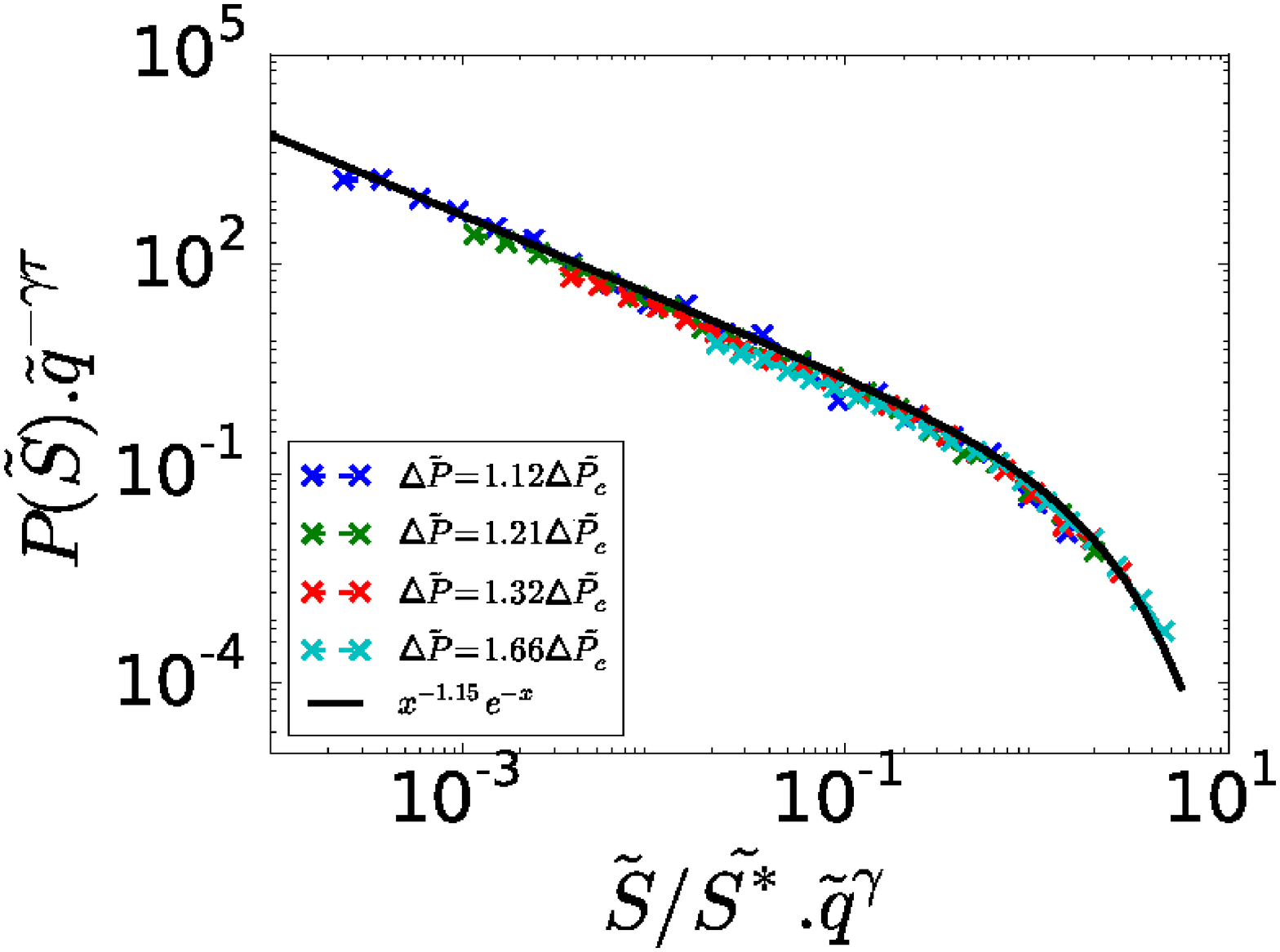}
	\includegraphics[width=0.3\linewidth]{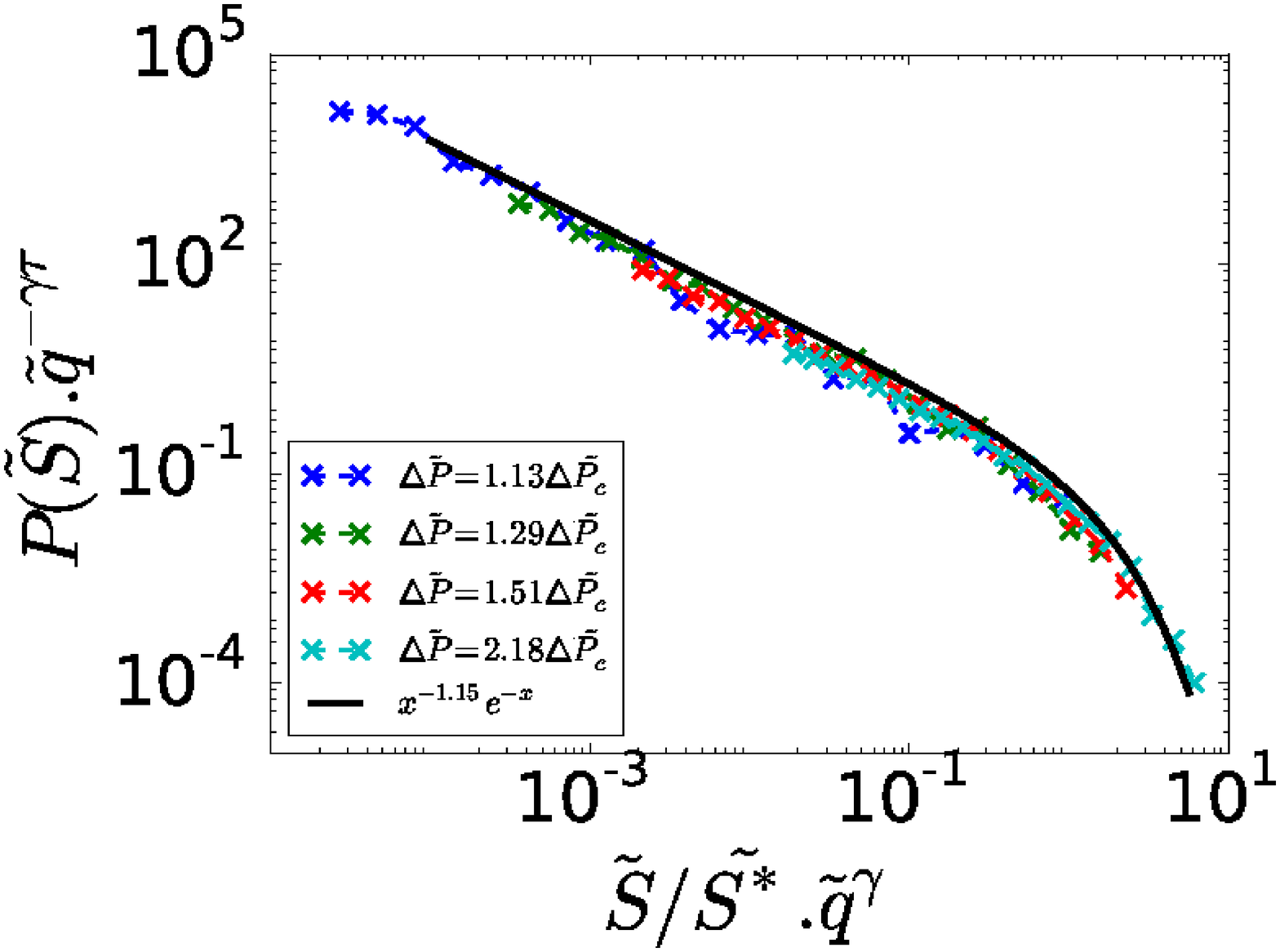}
	\caption{Renormalized cluster size probability distributions as function of the applied pressure for different $\std$: from left to right $\std = 0.5$, $2$ and $3$.\label{mult_sigma}. The continous line corresponds to a fit with eq. \eqref{eq:P_S}. $\tilde{S^*}$ is a renormalisation size proportional to $\sigma^2$: $\tilde{S^*}=3\sigma^2$.}
\end{figure*}

We first study the size distribution.
In  Fig. \ref{P_S} (left), we plot  the cluster size distribution $P(\tilde{S})$ for a given $\sigma=1$ at different applied pressure difference.
We can see that for any pressure difference, the distribution  follows  a power law for small sizes but with a cut-off at larger sizes.
The value of this cut-off size decreases with the pressure difference.
We then fit each distribution with the following law:
\begin{equation}
\label{eq:P_S}
	 P(\tilde{S})\propto \tilde{S}^{-\Spower}e^{-\tilde{S}/\aScut},
\end{equation}
where $\aScut$ is the characteristic cut-off size.
As it can be seen in the inset of Fig. \ref{P_S}, the cut-off size is found to follow a power law with the flow rate:
\begin{equation}
\label{eq:Scut_Q}
	\aScut \propto \tilde{q}^{-\Scutpower}.
\end{equation}

In Fig. \ref{P_S} (right), we plot the same distribution but using rescaled variables according to the eqs. \eqref{eq:P_S} and \eqref{eq:Scut_Q}. We can then see that all the data are collapsing on a master curve.
In addition, in Fig. \ref{mult_sigma}, we traced the rescaled distribution for 
different heterogeneity parameter $\sigma$.
As can be seen, the size of the clusters follows the same distribution function with the same exponents: $\Spower$ and $\Scutpower$ for any sigma.
We note, however, that the prefactor of the cut-off function varies with the disorder:
$\aScut = D(\sigma) \; \tilde{q}^{-\Scutpower}$, where $D(\std)=3 \sigma^2$.

Eqs. \eqref{eq:P_S} and \eqref{eq:Scut_Q} are thus reminiscent of what has been observed at the microscale.  The power law distribution characterizes the multi-scale nature of the problem.  At any pressure, cluster sizes are distributed within large range of sizes.
The higher bound, $\aScut$ decreases with the flow rate (or pressure).
Thus the closer we get to the global critical pressure, the wider the range of the power law distribution.
Qualitatively, this decrease can be understood by the fact that the non-flowing areas are divided into smaller ones with the appearance of new branches as the pressure increases.

As for the flow rate - pressure cure, it is worth mentioning that if the general trend is similar to the miscroscale, the main difference lies in the exponent value.
 Indeed, for  $\sigma=1$, we measure here $\Spower=1.15 \pm 0.05$ and $\Scutpower=1 \pm 0.05$. 
The $\Scutpower$ is thus similar the microscale  ($\gamma =  $ in \cite{chevalier15a})
 but the exponent $\Spower$ is significantly different ($\Spower = 1.5$ at microscale). This indicates again that the process of branching is physically different for the two scales because  the equations which are different (Stokes vs. Darcy) but also because the disorder  is different (correlated log-normal field vs. random cylinder packing).

\subsubsection{Cluster shape and Roughness}

\begin{figure}		
	\includegraphics[width=0.9\linewidth]{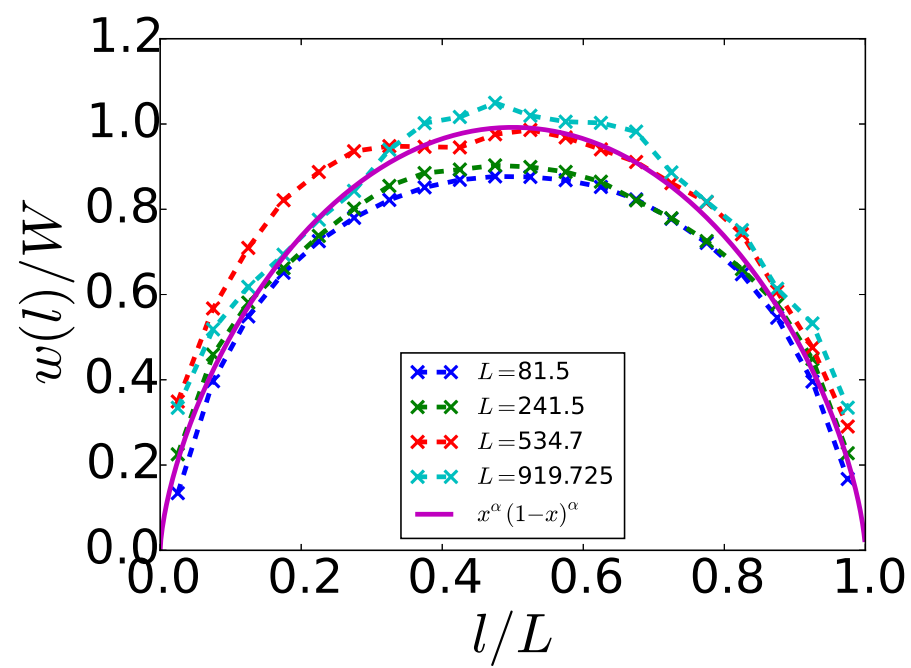}
	\caption{Averaged shape of the clusters  as function of their length. The continuous line corresponds to the fit witheq.  \eqref{eq:parabola}. \label{shape}
	}
\end{figure}

Now that we have shown that the cluster size is a multi-scale problem, we investigate the geometrical aspect of each cluster, as function of the applied pressure but also as function of the scale considered.
For a given length scale $\alength$, we  average the maximum width over clusters sharing the same length, $\langle \awidth \rangle_{\alength}$.
Moreover, we also study the shape of each cluster, $\tilde{w}(x)$,  defined as the local width as function of the relative $x$ coordinate. 
We also average this width function over many clusters sharing the same total length: 
$<\tilde{w}(x)>_{\alength}$.

In Figure \ref{shape}, we plot the contour shape $\langle \tilde{w}(x) \rangle_{\alength}$ for different total lengths $\alength$.
As expected, the cluster shape has a non-monotonic shape starting and ending with zero (\emph{e.g.} $\langle \tilde{w}(x=0) \rangle_{\alength}=\langle \tilde{w}(x=\alength) \rangle _{\alength}$).
The most interesting feature about the cluster shape is the fact that it is invariant to the  scale as shown in Fig. \ref{shape}.
In this figure, we renormalize the contour shape by the maximum width and the $x$ coordinate by the total length $\alength$.  All the contours then collapse on a parabolic-like master curve. This curve can be reasonably fitted to: 
\begin{equation}
\label{eq:parabola}
 B x^\rug (1-x)^{\rug},
 \end{equation} with $\rug = 0.75 \pm 0.05$.
We also observed that the renormalized clusters' shape (not shown) is also invariant with $\std$ and can be fitted with the same exponent.

\begin{figure}
	\includegraphics[width=0.9\linewidth]{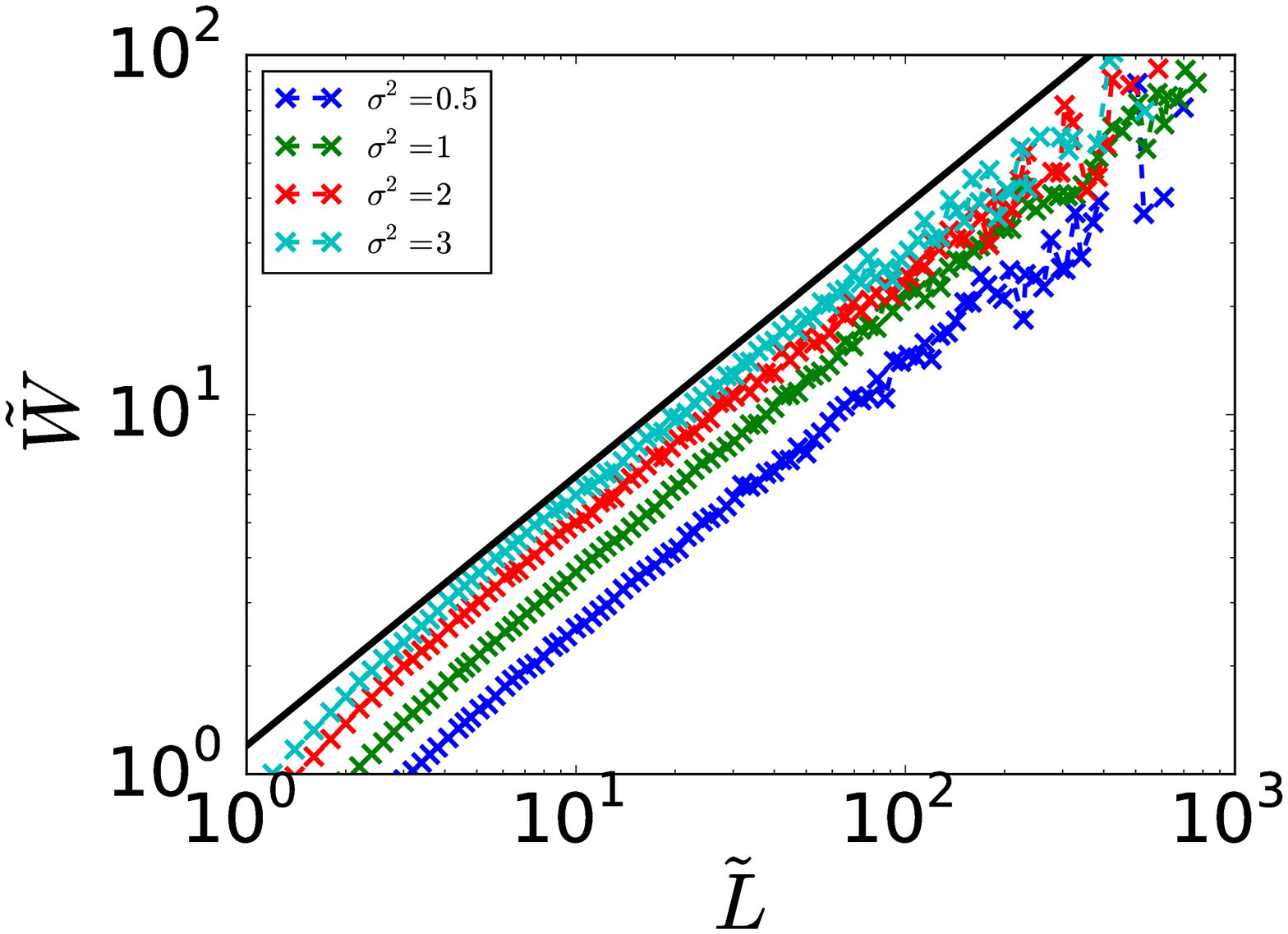}
	\caption{\label{W_L_mult_sigm} Mean width of the clusters as function of their length $\awidth(\alength)$ for different $\std$. The continuous line corresponds to a power-law with exponent $\rug=0.75$.}
\end{figure}

In addition, in Fig. \ref{W_L_mult_sigm}, we plot the maximum width according to the length of the cluster for different $\std$. We observe that the maximum width also follows a power law:
\begin{equation}
\label{eq:roughness}
	\alength \propto \awidth^\rug,
\end{equation}
with an exponent independent of the disorder.
The curve is, however, shifted with $\std$, reflecting the fact that the aspect ratio varies with $\std$. At low $\std$, the clusters are more elongated in the direction of flow, as can be clearly seen in the snapshot of the figure \ref{Velocity_map}.

These results demonstrate the fractal (self-affine) nature of the flow field.
At a given pressure, the size of the clusters is very widely distributed but with a shape which is invariant with the scale considered.

It is important to note that this behavior results from the self-affine roughness of the flowing paths surrounding the cluster.
For instance, if one would assume that a cluster is delimited by two paths from a random walk model, as proposed by \cite{kawagoe17} in another context, the cluster shape would then be described by  a first-time return random walk. 
It would thus follow relationships like  eqs. \eqref{eq:parabola} and \eqref{eq:roughness} but with an exponent $\rug=1/2$, which is the roughness of a random walk path.

Similarly as what has been observed earlier, the trend is comparable to the microscale. But here also, the main difference lies in the exponent value ($\rug =2/3 \pm 0.03$ at the pore scale) which is slightly higher.
At the microscopic scale, the exponent could be justified with the roughness of the directed polymer problem.

 We can put forward the following arguments to explain this small discrepancy.
First, due to numerical constraints, our statistical average is relatively low (ten realizations).
Secondly, the difference could also come from the log-normal distribution of the threshold, which might be considered as sufficiently "extreme" to deviate from the standard directed percolation.

It can also be argued that the paths are not necessarily directed in our problem. But, as it can be seen in the snapshots of Fig. \ref{Velocity_map}, for low $\std$,  overhangs  are no longer present but the roughness still remains the same.

\section{Discussion/Conclusion}

We have studied the flow of Bingham fluid in heterogeneous macroscopic porous media.
We have shown that at the macroscppic level (Darcy's scale) the flow properties share common features with the problem at the pore level.
First, due to the disorder, the mean flow rate - pressure curve evolves non-linearly over a certain pressure range.
Secondly, in this range, the velocity field exhibits multi-scale (fractal) properties that are reminiscent to other statistical physical problems with a critical transition (\emph{e.g} percolation, avalanches, etc.).
These multi-scale properties are described by power-law which are characterized by their exponent, $\flowpower$, $\Spower$, $\Scutpower$ and $\rug$.
Yet, a significant difference between the two scales lies in the value of these exponents.
At the macroscopic level, the mean flow rate increases with a power $\flowpower$ close to $3$ whereas at the microscale this exponent is close to $2$.
We attribute this difference to the fact that at the macroscopic level, each individual flow path widen as the pressure  increases. 
The exponents describing the flow geometry are also relatively different. At the macroscopic level, we observed $(\Spower, \Scutpower, \rug)= (1.15 \pm 0.05, 1 \pm 0.05, 0.75\pm 0.03)$, whereas at the microscopic level we had $(\Spower, \Scutpower, \rug)= (1.5 \pm 0.05, 1.1 \pm 0.05, 0.69\pm 0.03)$.
Thus, the distribution of sizes ($\Spower$) is  very different, the cluster roughness is slightly different and the evolution of the cut-off size with the flow rate is similar.
Another interesting results is the evolution of these scaling laws with the parameter of disorder $\std$.
Indeed, despite the fact that the flow field is at first sight very different when we vary $\std$, we observed that the exponents remain the same.
Only the prefactor of the different scaling laws are varying with $\std$.

We plan to extend this work in several directions. First, it would be interesting to investigate different types of permeability disorder. Even though  the log-normal distribution is a very popular model in geostatitics it would be interesting to investigate other distributions. 
Since the branching mechanism results from the balance between finding the most favorable regions and the length to reach them, it should depend on the statistics of the large permeability events. One could thus expect to observe different scaling laws if the distribution presents fat tails (\emph{e.g} power law).
Another direction could be to investigate the transport phenomena (temperature, chemical species, etc.) associated with this type of flow. Since the flow structure is very complex, one would expect a  non-linear transport properties depending on the applied  pressure.

\paragraph*{Acknowledgements}
This work benefits from a  PhD funding (R. Kostenko) from the French national radioactive waste management agency (Andra)

\bibliographystyle{elsarticle-num} 

\begin{thebibliography}{10}
\expandafter\ifx\csname url\endcsname\relax
  \def\url#1{\texttt{#1}}\fi
\expandafter\ifx\csname urlprefix\endcsname\relax\def\urlprefix{URL }\fi
\expandafter\ifx\csname href\endcsname\relax
  \def\href#1#2{#2} \def\path#1{#1}\fi

\bibitem{coussot05}
P.~Coussot, Rheometry of pastes, suspensions, and granular materials:
  applications in industry and environment, John Wiley and Sons, 2005.

\bibitem{entov67}
V.~Entov, On some two-dimensional problems of the theory of filtration with a
  limiting gradient., Prikl.~Mat.~Mekh. 31 (1967) 820--833.

\bibitem{prudhomme95}
R.~K. Prud'homme, Foams: Theory: Measurements: Applications, Vol.~57, CRC
  Press, 1995.

\bibitem{rossen90a}
W.~R. Rossen,
  \href{http://www.sciencedirect.com/science/article/pii/002197979090074X}{Theory
  of mobilization pressure gradient of flowing foams in porous media: I.
  incompressible foam}, J. Colloid Interface Sci. 136~(1) (1990) 1 -- 16.
\newblock \href {http://dx.doi.org/10.1016/0021-9797(90)90074-X}
  {\path{doi:10.1016/0021-9797(90)90074-X}}.
\newline\urlprefix\url{http://www.sciencedirect.com/science/article/pii/002197979090074X}

\bibitem{barbati16}
A.~C. Barbati, J.~Desroches, A.~Robisson, G.~H. McKinley,
  \href{https://doi.org/10.1146/annurev-chembioeng-080615-033630}{Complex
  fluids and hydraulic fracturing}, Annual Review of Chemical and Biomolecular
  Engineering 7~(1) (2016) 415--453, pMID: 27070765.
\newblock \href
  {http://arxiv.org/abs/https://doi.org/10.1146/annurev-chembioeng-080615-033630}
  {\path{arXiv:https://doi.org/10.1146/annurev-chembioeng-080615-033630}},
  \href {http://dx.doi.org/10.1146/annurev-chembioeng-080615-033630}
  {\path{doi:10.1146/annurev-chembioeng-080615-033630}}.
\newline\urlprefix\url{https://doi.org/10.1146/annurev-chembioeng-080615-033630}

\bibitem{al-fariss87}
T.~Al-Fariss, K.~L. Pinder,
  \href{http://dx.doi.org/10.1002/cjce.5450650306}{Flow through porous media of
  a shear-thinning liquid with yield stress}, Can. J. Chem. Eng. 65~(3) (1987)
  391--405.
\newblock \href {http://dx.doi.org/10.1002/cjce.5450650306}
  {\path{doi:10.1002/cjce.5450650306}}.
\newline\urlprefix\url{http://dx.doi.org/10.1002/cjce.5450650306}

\bibitem{chase05}
G.~Chase, P.~Dachavijit, A correlation for yield stress fluid flow through
  packed beds, Rheologica Acta 44~(5) (2005) 495--501.

\bibitem{chen05}
M.~Chen, W.~Rossen, Y.~C. Yortsos,
  \href{http://www.sciencedirect.com/science/article/pii/S0009250905001776}{The
  flow and displacement in porous media of fluids with yield stress}, Chem.
  Eng. Sci. 60~(15) (2005) 4183 -- 4202.
\newblock \href {http://dx.doi.org/DOI: 10.1016/j.ces.2005.02.054}
  {\path{doi:DOI: 10.1016/j.ces.2005.02.054}}.
\newline\urlprefix\url{http://www.sciencedirect.com/science/article/pii/S0009250905001776}

\bibitem{sochi08}
T.~Sochi, M.~J. Blunt,
  \href{http://www.sciencedirect.com/science/article/pii/S0920410507001106}{Pore-scale
  network modeling of ellis and herschel-bulkley fluids}, J. Pet. Sci. Eng.
  60~(2) (2008) 105 -- 124.
\newblock \href {http://dx.doi.org/10.1016/j.petrol.2007.05.009}
  {\path{doi:10.1016/j.petrol.2007.05.009}}.
\newline\urlprefix\url{http://www.sciencedirect.com/science/article/pii/S0920410507001106}

\bibitem{chevalier13}
T.~Chevalier, C.~Chevalier, X.~Clain, J.~Dupla, J.~Canou, S.~Rodts, P.~Coussot,
  \href{http://www.sciencedirect.com/science/article/pii/S0377025712002753}{Darcy's
  law for yield stress fluid flowing through a porous medium}, J. Non-Newtonian
  Fluid Mech. 195~(0) (2013) 57 -- 66.
\newblock \href {http://dx.doi.org/10.1016/j.jnnfm.2012.12.005}
  {\path{doi:10.1016/j.jnnfm.2012.12.005}}.
\newline\urlprefix\url{http://www.sciencedirect.com/science/article/pii/S0377025712002753}

\bibitem{talon13a}
L.~Talon, H.~Auradou, M.~Pessel, A.~Hansen,
  \href{http://stacks.iop.org/0295-5075/103/i=3/a=30003}{Geometry of optimal
  path hierarchies}, EPL (Europhysics Letters) 103~(3) (2013) 30003--.
\newline\urlprefix\url{http://stacks.iop.org/0295-5075/103/i=3/a=30003}

\bibitem{castro14}
A.~R. de~Castro, A.~Omari, A.~Ahmadi-S{\'e}nichault, D.~Bruneau,
  \href{http://dx.doi.org/10.1007/s11242-013-0248-5}{Toward a new method of
  porosimetry: Principles and experiments}, Transp. Porous Media 101~(3) (2014)
  349--364.
\newline\urlprefix\url{http://dx.doi.org/10.1007/s11242-013-0248-5}

\bibitem{nash16}
S.~Nash, D.~A.~S. Rees, The effect of microstructure on models for the flow of
  a bingham fluid in porous media, Transp. Porous Media.

\bibitem{shahsavari16}
S.~Shahsavari, G.~H. McKinley,
  \href{http://www.sciencedirect.com/science/article/pii/S0377025716300982}{Mobility
  and pore-scale fluid dynamics of rate-dependent yield-stress fluids flowing
  through fibrous porous media}, Journal of Non-Newtonian Fluid Mechanics 235
  (2016) 76 -- 82.
\newblock \href
  {http://dx.doi.org/http://dx.doi.org/10.1016/j.jnnfm.2016.07.006}
  {\path{doi:http://dx.doi.org/10.1016/j.jnnfm.2016.07.006}}.
\newline\urlprefix\url{http://www.sciencedirect.com/science/article/pii/S0377025716300982}

\bibitem{pascal81}
H.~Pascal, \href{https://doi.org/10.1007/BF01170343}{Nonsteady flow through
  porous media in the presence of a threshold gradient}, Acta Mechanica 39~(3)
  (1981) 207--224.
\newblock \href {http://dx.doi.org/10.1007/BF01170343}
  {\path{doi:10.1007/BF01170343}}.
\newline\urlprefix\url{https://doi.org/10.1007/BF01170343}

\bibitem{wu90}
Y.~Wu, K.~Pruess, P.~Witherspoon, Flow and displacement of bingham
  non-newtonian fluids in porous media.

\bibitem{rees15}
D.~A.~S. Rees, A.~P. Bassom,
  \href{http://www.sciencedirect.com/science/article/pii/S0017931014009326}{Unsteady
  thermal boundary layer flows of a bingham fluid in a porous medium},
  International Journal of Heat and Mass Transfer 82 (2015) 460 -- 467.
\newblock \href
  {http://dx.doi.org/https://doi.org/10.1016/j.ijheatmasstransfer.2014.10.047}
  {\path{doi:https://doi.org/10.1016/j.ijheatmasstransfer.2014.10.047}}.
\newline\urlprefix\url{http://www.sciencedirect.com/science/article/pii/S0017931014009326}

\bibitem{chevalier15a}
T.~Chevalier, L.~Talon,
  \href{http://link.aps.org/doi/10.1103/PhysRevE.91.023011}{Generalization of
  {D}arcy's law for {B}ingham fluids in porous media: From flow-field
  statistics to the flow-rate regimes}, Phys. Rev. E 91 (2015) 023011.
\newblock \href {http://dx.doi.org/10.1103/PhysRevE.91.023011}
  {\path{doi:10.1103/PhysRevE.91.023011}}.
\newline\urlprefix\url{http://link.aps.org/doi/10.1103/PhysRevE.91.023011}

\bibitem{chevalier15c}
T.~Chevalier, L.~Talon, Moving line model and avalanche statistics of {B}ingham
  fluid flow in porous media, Eur. Phys. J. E. 38 (2015) 76.
\newblock \href {http://dx.doi.org/10.1140/epje/i2015-15076-5}
  {\path{doi:10.1140/epje/i2015-15076-5}}.

\bibitem{stauffer91}
D.~Stauffer, A.~Aharony, Introduction to percolation theory, Taylor and
  Francis, 1991.

\bibitem{amaral95}
L.~A.~N. Amaral, A.~L. Barabasi, S.~V. Buldyrev, S.~T. Harrington, S.~Havlin,
  R.~Sadrlahijany, H.~E. Stanley, Avalanches and the directed percolation
  depinning model - experiments, simulations, and theory, Phys. Rev. E 51~(5)
  (1995) 4655--4673.
\newblock \href {http://dx.doi.org/10.1103/PhysRevE.51.4655}
  {\path{doi:10.1103/PhysRevE.51.4655}}.

\bibitem{chevalier17}
T.~Chevalier, A.~K. Dubey, S.~Atis, A.~Rosso, D.~Salin, L.~Talon,
  \href{https://link.aps.org/doi/10.1103/PhysRevE.95.042210}{Avalanches
  dynamics in reaction fronts in disordered flows}, Phys. Rev. E 95 (2017)
  042210.
\newblock \href {http://dx.doi.org/10.1103/PhysRevE.95.042210}
  {\path{doi:10.1103/PhysRevE.95.042210}}.
\newline\urlprefix\url{https://link.aps.org/doi/10.1103/PhysRevE.95.042210}

\bibitem{hewitt16}
D.~R. Hewitt, M.~Daneshi, N.~J. Balmforth, D.~M. Martinez, Obstructed and
  channelized viscoplastic flow in a hele-shaw cell, Journal of Fluid Mechanics
  790 (2016) 173–204.

\bibitem{hirasaki74}
G.~Hirasaki, G.~Pope, et~al., Analysis of factors influencing mobility and
  adsorption in the flow of polymer solution through porous media, Society of
  Petroleum Engineers Journal 14~(04) (1974) 337--346.

\bibitem{chauveteau82}
G.~Chauveteau, \href{https://doi.org/10.1122/1.549660}{Rodlike polymer solution
  flow through fine pores: Influence of pore size on rheological behavior},
  Journal of Rheology 26~(2) (1982) 111--142.
\newblock \href {http://arxiv.org/abs/https://doi.org/10.1122/1.549660}
  {\path{arXiv:https://doi.org/10.1122/1.549660}}, \href
  {http://dx.doi.org/10.1122/1.549660} {\path{doi:10.1122/1.549660}}.
\newline\urlprefix\url{https://doi.org/10.1122/1.549660}

\bibitem{papanastasiou87}
T.~C. Papanastasiou, \href{http://link.aip.org/link/?JOR/31/385/1}{Flows of
  materials with yield}, Journal of Rheology 31~(5) (1987) 385--404.
\newblock \href {http://dx.doi.org/10.1122/1.549926}
  {\path{doi:10.1122/1.549926}}.
\newline\urlprefix\url{http://link.aip.org/link/?JOR/31/385/1}

\bibitem{brinkman47}
H.~Brinkman, {A} calculation of the viscous forces exerted by a flowing fluid
  on a dense swarm of particles, Appl. Sci. Res. sect A1 (1947) 27--39.

\bibitem{gelhar83}
L.~Gelhar, C.~Axness, {T}hree-{D}imensional {S}tochastic {A}nalysis of
  {M}acrodispersion in {A}quifers, Water Resour. Res. 19 (1983) 161--180.

\bibitem{dagan82}
G.~Dagan, Stochastic modeling of groundwater flow by unconditional and
  conditional probabilities. 2. {T}he solute transport, Water Resour. Res. 18
  (1982) 835--848.

\bibitem{yiotis13}
A.~G. Yiotis, L.~Talon, D.~Salin,
  \href{http://link.aps.org/doi/10.1103/PhysRevE.87.033001}{Blob population
  dynamics during immiscible two-phase flows in reconstructed porous media},
  Phys. Rev. E 87 (2013) 033001.
\newblock \href {http://dx.doi.org/10.1103/PhysRevE.87.033001}
  {\path{doi:10.1103/PhysRevE.87.033001}}.
\newline\urlprefix\url{http://link.aps.org/doi/10.1103/PhysRevE.87.033001}

\bibitem{bird87}
R.~Bird, R.~Armstrong, O.~Hassager,
  \href{http://www.osti.gov/scitech/servlets/purl/6164599}{Dynamics of
  polymeric liquids. Vol. 1, 2nd Ed. : Fluid mechanics}, John Wiley and Sons
  Inc., New York, NY, 1987.
\newline\urlprefix\url{http://www.osti.gov/scitech/servlets/purl/6164599}

\bibitem{ginzburg15}
I.~Ginzburg, G.~Silva, L.~Talon,
  \href{http://link.aps.org/doi/10.1103/PhysRevE.91.023307}{Analysis and
  improvement of brinkman lattice boltzmann schemes: Bulk, boundary, interface.
  similarity and distinctness with finite elements in heterogeneous porous
  media}, Phys. Rev. E 91 (2015) 023307.
\newblock \href {http://dx.doi.org/10.1103/PhysRevE.91.023307}
  {\path{doi:10.1103/PhysRevE.91.023307}}.
\newline\urlprefix\url{http://link.aps.org/doi/10.1103/PhysRevE.91.023307}

\bibitem{ginzburg17}
I.~Ginzburg,
  \href{http://link.aps.org/doi/10.1103/PhysRevE.95.013304}{Prediction of the
  moments in advection-diffusion lattice boltzmann method. i. truncation
  dispersion, skewness, and kurtosis}, Phys. Rev. E 95 (2017) 013304.
\newblock \href {http://dx.doi.org/10.1103/PhysRevE.95.013304}
  {\path{doi:10.1103/PhysRevE.95.013304}}.
\newline\urlprefix\url{http://link.aps.org/doi/10.1103/PhysRevE.95.013304}

\bibitem{roux87}
S.~Roux, H.~J. Herrmann,
  \href{http://stacks.iop.org/0295-5075/4/i=11/a=003}{Disorder-induced
  nonlinear conductivity}, Europhys. Lett. 4~(11) (1987) 1227.
\newline\urlprefix\url{http://stacks.iop.org/0295-5075/4/i=11/a=003}

\bibitem{kardar87}
M.~Kardar, Y.-C. Zhang,
  \href{http://link.aps.org/doi/10.1103/PhysRevLett.58.2087}{Scaling of
  directed polymers in random media}, Phys. Rev. Lett. 58 (1987) 2087--2090.
\newblock \href {http://dx.doi.org/10.1103/PhysRevLett.58.2087}
  {\path{doi:10.1103/PhysRevLett.58.2087}}.
\newline\urlprefix\url{http://link.aps.org/doi/10.1103/PhysRevLett.58.2087}

\bibitem{halpin-healy95}
T.~Halpin-Healy, Y.-C. Zhang,
  \href{http://www.sciencedirect.com/science/article/pii/037015739400087J}{Kinetic
  roughening phenomena, stochastic growth, directed polymers and all that.
  aspects of multidisciplinary statistical mechanics}, Physics Reports 254~(4)
  (1995) 215 -- 414.
\newblock \href
  {http://dx.doi.org/https://doi.org/10.1016/0370-1573(94)00087-J}
  {\path{doi:https://doi.org/10.1016/0370-1573(94)00087-J}}.
\newline\urlprefix\url{http://www.sciencedirect.com/science/article/pii/037015739400087J}

\bibitem{gueudre15}
T.~Gueudre, P.~Le~Doussal, J.-P. Bouchaud, A.~Rosso,
  \href{https://link.aps.org/doi/10.1103/PhysRevE.91.062110}{Ground-state
  statistics of directed polymers with heavy-tailed disorder}, Phys. Rev. E 91
  (2015) 062110.
\newblock \href {http://dx.doi.org/10.1103/PhysRevE.91.062110}
  {\path{doi:10.1103/PhysRevE.91.062110}}.
\newline\urlprefix\url{https://link.aps.org/doi/10.1103/PhysRevE.91.062110}

\bibitem{kawagoe17}
K.~Kawagoe, G.~Huber, M.~Pradas, M.~Wilkinson, A.~Pumir, E.~Ben-Naim,
  \href{https://link.aps.org/doi/10.1103/PhysRevE.96.012142}{Aggregation-fragmentation-diffusion
  model for trail dynamics}, Phys. Rev. E 96 (2017) 012142.
\newblock \href {http://dx.doi.org/10.1103/PhysRevE.96.012142}
  {\path{doi:10.1103/PhysRevE.96.012142}}.
\newline\urlprefix\url{https://link.aps.org/doi/10.1103/PhysRevE.96.012142}

\end{thebibliography}

\end{document}